\documentclass[showpacs,preprintnumbers,amsmath,amssymb,superscriptaddress,nofootinbib,english]{revtex4}

\usepackage{graphicx,floatflt}
\usepackage{amsmath}
\usepackage{amssymb}

\usepackage{epsfig}
\usepackage{color}

\newcommand{\dd}{\mathrm{d}} 
\newcommand{\Mpl}{M_\mathrm{pl}} 

\newcommand{\epsone}{\epsilon_1}
\newcommand{\epstwo}{\epsilon_2}
\newcommand{\rr}{\mathrm}
\newcommand{\ns}{n_{\rr s} }

\newcommand{\phic}{\phi_{\rr c}}

\newcommand{\Nend}{N_{\rr {end}}}

\newcommand{\fNL}{f_{\mathrm{NL}}}

\newcommand{\Mpc}{\rr{Mpc}}
\newcommand{\xend}{x_{\rr{end}}}
\newcommand{\kp}{k_{\rr{p}}}
\newcommand{\kd}{k_{\rr{d}}}

\newcommand{\be}{\begin{equation}}
\newcommand{\ee}{\end{equation}}
\newcommand{\ba}{\begin{align}}
\newcommand{\ea}{\end{align}}

\newcommand{\lsim}
{\;\raisebox{-.3em}{$\stackrel{\displaystyle <}{\sim}$}\;}
\newcommand{\gsim}
{\;\raisebox{-.3em}{$\stackrel{\displaystyle >}{\sim}$}\;}

\newcommand{\idistort}{\texttt{idistort }}
\newcommand{\Greens}{\texttt{Greens }}

\begin{document}

\preprint{TUM-HEP-931-14}

\title{Testing Inflation and Curvaton Scenarios with CMB Distortions}

\author{S\'ebastien Clesse} \email{s.clesse@tum.de}
\email[]{sebastien.clesse@unamur.be}
\affiliation{Namur Center of Complex Systems (naXys), Department of Mathematics, University of Namur, Rempart de la Vierge 8, 5000 Namur, Belgium}
\author{Bj\"orn Garbrecht} \email{garbrecht@tum.de} 
\affiliation{Physik Department T70, James-Franck-Stra{\ss}e,
Technische Universit\"at M\"unchen, 85748 Garching, Germany}
\author{Yi Zhu} \email{yi.zhu@tum.de}
\affiliation{Physik Department T70, James-Franck-Stra{\ss}e,
Technische Universit\"at M\"unchen, 85748 Garching, Germany}
\date{\today}

\begin{abstract}

Prior to recombination, Silk damping causes the dissipation of energy from acoustic waves into the monopole of the Cosmic Microwave Background (CMB), resulting in spectral distortions. These can
be used to probe the primordial scalar power spectrum on smaller scales than it is possible with CMB anisotropies.  An enhancement of power on these scales is nevertheless required for the resulting distortions to be detectable by future experiments like PIXIE.  In this paper, we examine all 49 single-field inflation models listed by Martin et al. in the \textit{Encyclopaedia Inflationaris}~\cite{Martin:2013tda} and find that only one of these may lead to a detectable level of distortions in a tuned region of its parameter space, namely the original hybrid model.   Three effective multi-field scenarios are also studied: with softly and suddenly turning trajectories, and with a mild waterfall trajectory. Softly turning trajectories do not induce  distortions at any detectable level, whereas a sudden turn in the field space or a mild 
waterfall trajectory predicts a peak (plus damped oscillations in the sudden turn case) in the scalar power spectrum, which can lead to an observable amount of CMB distortions.  Finally, another scenario leading to potentially detectable distortions involves a curvaton whose blue spectrum is subdominant on CMB angular scales and overtakes the inflaton spectrum on smaller scales. In this case however, we show that the bounds from ultra compact minihaloes are not satisfied.  Expectations for an ultimate PRISM-class experiment characterized by an
improvement in sensitivity by a factor of ten are discussed for some models. 

\end{abstract}
\maketitle

\section{Introduction}

During the last two decades, observations of the Cosmic Microwave Background (CMB) anisotropies have been allowing for an increasingly accurate determination of the energy content and the shape of the Universe, as well as of the primordial density perturbations, that are at the origin of large scale structure formation~\cite{Ade:2013lta}.  In order to match observations, these primordial inhomogeneities can be described as Gaussian and have a nearly scale-invariant power spectrum, which is a generic prediction of inflation models in which the Universe undergoes an early phase of exponentially accelerated expansion.   The most recent experiments like the Atacama Cosmology Telescope~\cite{Das:2013zf}, and the South Pole Telescope~\cite{Story:2012wx} have measured the CMB angular power spectrum up to multipoles $l=l_{\rm max} \sim 3000$, while Planck~\citep{Planck:2013kta} observes at the same time large scales
up to the lowest multipoles as well.  This range of angular scales probes wavelength modes of primordial density perturbations within the window $ k_{l \rm max}=2 \times 10^{-4} \Mpc^{-1} \lesssim k \lesssim 0.2 \Mpc^{-1}=k_{l \rm min}$.  In the context of inflation, the largest have exited the Hubble radius about 60-efolds before the end of inflation and have since been frozen until they recently have re-entered the horizon.  CMB anisotropy experiments can therefore probe about $7\approx\ln  k_{\ell \rm max}/k_{\ell \rm min}$ e-folds of inflationary expansion.  Earlier and later periods of the inflationary epoch remain inaccessible to observations, respectively either because the corresponding scales are still super-horizon today, or because they have led to acoustic oscillations that are highly suppressed due to Silk damping.   Recently the BICEP2~\cite{Ade:2014xna} experiment has claimed the discovery of B-mode polarization in the CMB spectrum induced by primordial gravitational waves corresponding to a tensor to scalar ratio $r = 0.20 ^{+0.07}_{-0.05} $.  These results are nevertheless still controversial (see e.g.~\cite{Mortonson:2014bja}) and thus we adopt a conservative approach and do not consider BICEP2 constraints on the studied models in this paper. 

Other signals and observational techniques must therefore be envisaged in order to extend the available information on the inflationary epoch by several e-folds. One example are the 21cm angular power spectra from the dark ages and the subsequent reionisation (see e.g Refs.~\cite{Pritchard:2011xb,Pritchard:2010pa,Furlanetto:2009qk,Clesse:2012th}).  Measuring distortions of the CMB black-body radiation is another, less futuristic, interesting possibility.  Spectral distortions can be the consequence of energy injection into the primeval plasma through several processes~\cite{Sunyaev:2013aoa,Chluba:2013wsa,Chluba:2011hw}: the decay or the annihilation of relic particles, the evaporation of primordial black holes, the interaction of CMB photons with non-relativistic electrons and baryons after recombination~\cite{Chluba:2011hw}, as well as recombination lines, reionisation and structure formation.  
Here, we are mainly interested in the energy stored in sound waves 
and that is dissipated due to Silk damping into the monopole,
which creates specific distortions of the CMB black-body spectrum~\cite{Chluba:2013dna,Khatri:2012rt,Chluba:2012we,Chluba:2012gq,Khatri:2013dha}.  Measuring such distortions can therefore in principle provide some information on the primordial power spectrum of density perturbations, on scales much smaller than the ones probed by CMB anisotropy experiments.  Future dedicated CMB distortion experiments could probe modes in the range $8 \lesssim k \lesssim 10^4 \Mpc^{-1}$, potentially extending the period of inflation that is accessible to observations up to 17 e-folds~\cite{Khatri:2013xwa}.

Any perturbations of the thermodynamic equilibrium between photons and baryons during the tightly coupled regime and towards its end can induce CMB spectral distortions.  At very early times (redshifts $z\gtrsim 2 \times 10^6$), thermalization is very efficient and erases distortions caused by any amount of energy injection.  However, at lower redshifts, spectral distortions may survive and still be observable today.  It is common practice to distinguish between distortions characterised by a frequency-dependent chemical potential $\mu(\nu)$, the so-called $\mu$-distortions, 
and distortions of the $y$-type mostly associated to SZ clusters.  For completeness, one one should also consider partially Comptonized intermediate $i$-distortions, as discussed in Refs.~\cite{Khatri:2012tw,Khatri:2013dha}.   
The current 95$\%$C.L. limits on $\mu$- and $y$-distortions result from the COBE-FIRAS experiment~\cite{Fixsen:1996nj} and are $y<1.5 \times 10^{-5}$ and $\mu < 9 \times 10^{-5}$.  Recently, the Primordial Inflation Explorer (PIXIE) \cite{Kogut:2011xw} experiment has been proposed. It is one of its objectives to improve these limits by about three orders of magnitude.    Further in the future, the next to next generation of experiments, of the class of the Polarized Radiation Imaging and Spectroscopy Mission (PRISM) \cite{Andre:2013nfa,Andre:2013afa}, will have the ability to improve the sensitivity to spectral distortions by about one order of magnitude compared to PIXIE.  

Some work has been done recently to forecast possible constraints on the amplitude of the power spectrum of curvature perturbations, its spectral index and the running of the latter, for various experimental configurations~\cite{Chluba:2012we,Khatri:2013dha,Chluba:2013dna}.  As a result, it is not expected that the PIXIE experiment, by probing only spectral distortions, will reach the sensitivity to improve current constraints on the spectral index in models for which its value on usual CMB scales can be extended to scales probed by CMB distortions.  But it is still unclear whether new constraints will be established on specific inflation models that predict an increase of power on scales smaller than the ones probed by CMB anisotropy experiments. It has also not yet been studied whether a curvaton, which is an alternative to the inflaton as the field responsible for the generation of the curvature
power spectrum, can lead to observable CMB distortions. 

In this paper, we adopt a model-oriented approach.  We have studied all the single-field inflation models listed in Ref.~\cite{Martin:2013tda} and have established a short list of models predicting an increase of power on small scales for some part of their parameter space. For each of these we have studied whether they can lead to an observable level of CMB distortions or not, and we have determined the regions in the parameter space where this occurs.  For this purpose, we have modified the \idistort template~\cite{Khatri:2012tw},  as well as the \Greens package~\cite{Chluba:2013vsa} for Markov-Chain-Monte-Carlo (MCMC) parameter estimation, in order to include any shape of the power spectrum of curvature perturbations, and for all the relevant models we have calculated the spectrum of CMB distortions.  In addition to single field models, we have considered three types of multi-field scenarios: i) an effective model where the field trajectories perform a turn with a constant angular velocity at some point of the inflationary evolution, ii) an effective model where trajectories perform a sudden turn in the field space, and iii) an 
effective hybrid model for which between $30$ and $60$ e-folds of expansion are realised 
during the classical field evolution in the final waterfall phase.  Finally, we study a curvaton scenario that can possibly induce a detectable level of distortions.  We explore the signatures of each model on the CMB black-body spectrum and determine whether they could be measured by PIXIE.  For some models we also discuss expectations for a PRISM-class experiment. 

The paper is organized as follows: In Sec.~\ref{sec:distintro}, the different types of distortions are introduced.  Sec.~\ref{sec:sensitivity} is dedicated to the expected sensitivity of the PIXIE experiment  and to forecasts on distinguishing different shapes of the scalar power spectrum.  In Sec.~\ref{sec:single_field}, single field models that possibly lead to CMB distortions observable by PIXIE are studied.   We investigate three effective scenarios of multi-field inflation in Sec.~\ref{sec:multi_field} and focus on a specific curvaton model in Sec.~\ref{sec:curvaton}. To exemplify the benefit of improving the spectrometer's sensitivity beyond the level proposed for PIXIE, in Sec.~\ref{sec:PRISM}, we investigate how a PRISM-class probe could be used to determine the running of the spectral index or a possible sub-dominant
contamination of the scalar power spectrum from a curvaton. Finally, we summarize our findings and discuss the possible implications for the dedicated CMB distortion experiments of the next generation in the Conclusions (Sec.~\ref{sec:ccl}).

\section{$\mu$-type, $y$-type and intermediate $i$-type CMB distortions}  \label{sec:distintro}

\subsection{The thermal SZ effect}

In the deep gravitational wells formed by clusters of galaxies, despite the fact that photons outnumber baryons and electrons, the electrons acquire sufficiently large thermal velocities (the electron temperature exceeds the photon temperature by more than eight orders of magnitudes) to Comptonize the CMB photons.  This phenomenon is called the \textit{thermal SZ effect}~\cite{SZthermal} and induces a change in the intensity of the CMB black-body radiation given by
\be \label{eq:deltaI_ydist}
\Delta I (\nu) = y \frac{x \rr e^ x}{\rr e^ x - 1} \left[\frac{x (\rr e^ x + 1)}{\rr e^ x - 1} - 4  \right] I_0 (\nu)\,,
\ee
where $x \equiv h \nu / (k_B T_0)$, $\nu$ is the frequency, $k_B$, $h$ the Boltzmann and Planck constants,  $T_0 $ is the CMB black-body temperature and $I_0 (\nu)$ is its intensity.  The key parameter $y$ is given by the line of sight integral
\be
y = \int \frac{k_B T_e n_e \sigma_T a^ 2 H}{m_e c^ 2} \dd \eta\,,
\ee
where $m_e$, $T_e$ and $n_e$ are respectively the electron mass, temperature and number density, $\sigma_T$ is the Thomson cross-section and $\eta$ the conformal time.  Typical values for galaxy clusters are $y\sim 10^{-4}$.  Such $y$-distortions correspond to a reduction of intensity of the CMB spectrum at frequencies where the Rayleigh-Jeans limit is valid and to an increase of intensity in the Wien part of the spectrum.  Note that distortions that bear
the form of Eq.~(\ref{eq:deltaI_ydist}) but which are not due to the thermal SZ effect are referred to as $y$-distortions as well. 

\subsection{Before recombination: distortions induced by energy release}

The energy spectrum of the CMB gives us some information about the thermal history of the Universe and the physical processes taking place during the tight-coupling regime of radiation and baryonic matter and towards its end.

Prior to recombination, the thermal equilibrium can be broken due to energy injection, even if electrons and ions remain in thermal equilibrium due to Coulomb collisions.  One can distinguish four phases, which we describe in the short overview below.  We refer to Ref.~\cite{Chluba:2011hw} and references therein for more a detailed discussion.
\begin{enumerate}
\item At $z \gtrsim 2 \times 10^6$:  The CMB corresponds to an almost perfect black body even in the presence of a large energy injection.  Thermal equilibrium between photons, electrons and baryons is maintained via Compton scattering, which quickly redistributes any excess or deficit of energy over the entire spectrum of photons.  Double Compton scattering and photon production by Bremsstrahlung drives the chemical potential to zero.  Compton scattering with photons whose number density is $\sim 10^9$ times the number density of electrons also maintains the Maxwellian distribution of electrons.  

\item At $2 \times 10^5 \lesssim z \lesssim 2 \times 10^6$:  Kinetic equilibrium is maintained by Compton scattering, but the double Compton scattering and the Bremsstrahlung processes are inefficient in creating photons.  If for some reason (decay/annihilation of relic particles, dissipating acoustic waves, evaporation of primordial black holes) energy is injected in the plasma, then the deviations from the black body can be described by a Bose-Einstein distribution with a chemical potential $\mu$. CMB distortions generated during this
epoch are therefore of the $\mu$-type.  	

\item At $1.5 \times 10^4 \lesssim z  \lesssim 2 \times 10^5 $ : In the presence of energy injection, the Bose-Einstein distribution of the photons cannot be maintained any more by Compton scattering.  Kinetic equilibrium with electrons and can only be established partially, and the CMB spectrum is partially Comptonized.  The distortions induced in the CMB black body spectrum are thus of the intermediate $i-$type.

\item At $z \lesssim 1.5 \times 10^4$:  The efficiency of Compton scattering is minimal and kinetic equilibrium with electrons cannot be established. Energy injection induces deviations to the Planckian CMB spectrum in the form of $y-$type distortions.    

\end{enumerate}
The different types of distortions can be used to discriminate between models and energy injection processes.  

\subsection{Distortions from dissipating acoustic waves in the Silk-damping tail}
  
After re-entering inside the horizon, density perturbations generate acoustic waves which cannot propagate for perturbation modes whose wavelength is smaller than the mean free path of photons.  This induces the damping of the acoustic peaks in the tail of the CMB angular power spectrum.  The energy initially stored within these acoustic oscillations is therefore dissipated into the monopole.  

One can define the fractional dissipated energy $Q \equiv \Delta E /E$. In the tight coupling regime, the evolution equation for $Q$ is given by~\cite{Chluba:2012gq}
\be
\frac{\dd Q}{\dd t} \simeq \frac{9}{4} \frac{\dd (1/k_{\rr D}^2)}{\dd z} \int \frac{\dd k^3}{(2 \pi)^3} k^2 P^\gamma_i (k) \rr e^{-k^2/k_{\rr D}^2}\,,
\ee
where $P_i^\gamma (k) \equiv 4 P_\zeta (k) / (0.4 R_\nu +1.5) $ ($Q$ is therefore related to the power spectrum of curvature perturbations $P_\zeta$) and $R_\nu = \rho_\nu /(\rho_\nu + \rho_\gamma) \simeq 0.4$, and where $k_{\rr D}$ is the Silk damping scale.  
If one wants to evaluate the total modifications of the CMB spectrum induced by energy injection described by the function $Q(t)$, it is necessary to solve the Kompaneets equation numerically.  For this purpose, the \idistort template has been developed~\cite{Khatri:2012tw}, which we employ in the present work.   For standard power spectrum parameter estimation with a MCMC method we have used a modified version of the \Greens code~\cite{Chluba:2013vsa}, which is based on a Green's function approach to approximate efficiently the spectral distortion spectra. 

\section{Sensitivity of the PIXIE experiment} \label{sec:sensitivity}

 The Primordial Inflation Explorer (PIXIE) experiment~\cite{Kogut:2011xw} has been proposed recently as a NASA Explorer class mission in order to map the absolute intensity and the linear polarization of the CMB as well as to measure its absolute frequency spectrum.   One major objective is to constrain the ratio $r$ between primordial tensor and scalar perturbations by measuring the polarization of the CMB on large angular scales. Another important scientific goal is the measurement of the spectral distortions of the CMB black body spectrum over a wide range of frequencies.  

PIXIE will probe frequencies from 30~GHz to 6~THz, in 400 spectral channels of 15~GHz bandwidths.  The instrument sensitivity to the unpolarized signal in each frequency bin will be
\be
\label{sensitivity:PIXIE}
\delta I_\nu ^{\rr{PIXIE}} = 5 \times 10^{-26} \rr{W m^{-2} Sr^{-1} Hz^{-1}}~,
\ee 
which will allow a dectection of
distortions characterised by~\cite{Kogut:2011xw}
\be
\label{PIXIE:mu:5sigma}
\mu = 5 \times 10^{-8}\;\; \textnormal{and}\;\; y = 1 \times 10^{-8}\;\;\textnormal{at}\;\;5\sigma\,.
\ee

\begin{figure}
\begin{center}
\includegraphics[width=10cm]{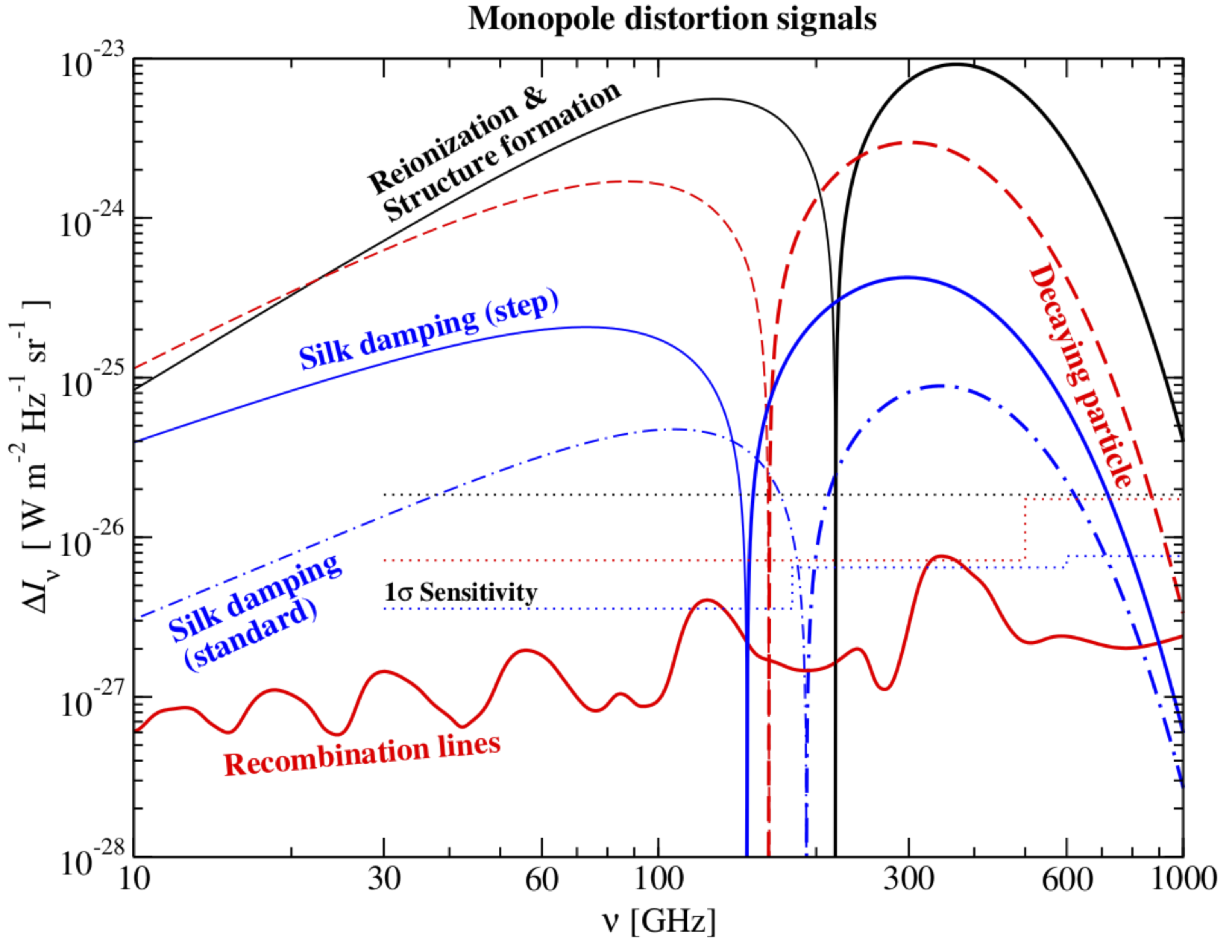}
\caption{ \label{fig:gendist} Spectral distortions from various sources and $1\sigma$ sensitivity for PRISM (dotted) for different designs.   The most important source comes from $y-$type distortions from reionization and structure formation (solid black line).  Scenarios such as decaying dark matter particles may be another important source of distortions (dashed red line).  Compared to Silk damping for a standard inflation scenario, with no important running (dashed-dot blue line), more complicated scenarios (e.g. step in the scalar power spectrum, solid blue line) might produce a larger distortion signal.  Recombination lines generate a spectral distortions with a lower amplitude (solid red line). Thin lines denote a negative distortion signal whereas tick lines denote a positive signal. Figure from Ref.~\cite{Andre:2013nfa}, where the parametric details are specified.}
\end{center}
\end{figure}

For comparison, reionisation and structure formation predict distortions with a maximum of $\delta I_\nu = 10^{-23} \rr{W m^{-2} Sr^{-1} Hz^{-1}} $, standard single-field slow-roll models (with no running) $\delta I_\nu \sim 10^{-25} \rr{W m^{-2} Sr^{-1} Hz^{-1}} $ whereas distortions due to recombination lines do not exceed $\delta I_\mu \sim 10^{-26} \rr{W m^{-2} Sr^{-1} Hz^{-1}}  $.  Some typical distortion spectra are displayed in FIG.~\ref{fig:gendist}.  One can clearly see from that figure that for frequencies close to $\nu \simeq 200 \rr{GHz}$, there exist a window where distortions due to reionisation and structure formation vanish.  This is the window that may be used for the detection of intermediate type distortions from inflation or from decaying particles.  

\begin{figure}
\begin{center}
\includegraphics[width=16cm]{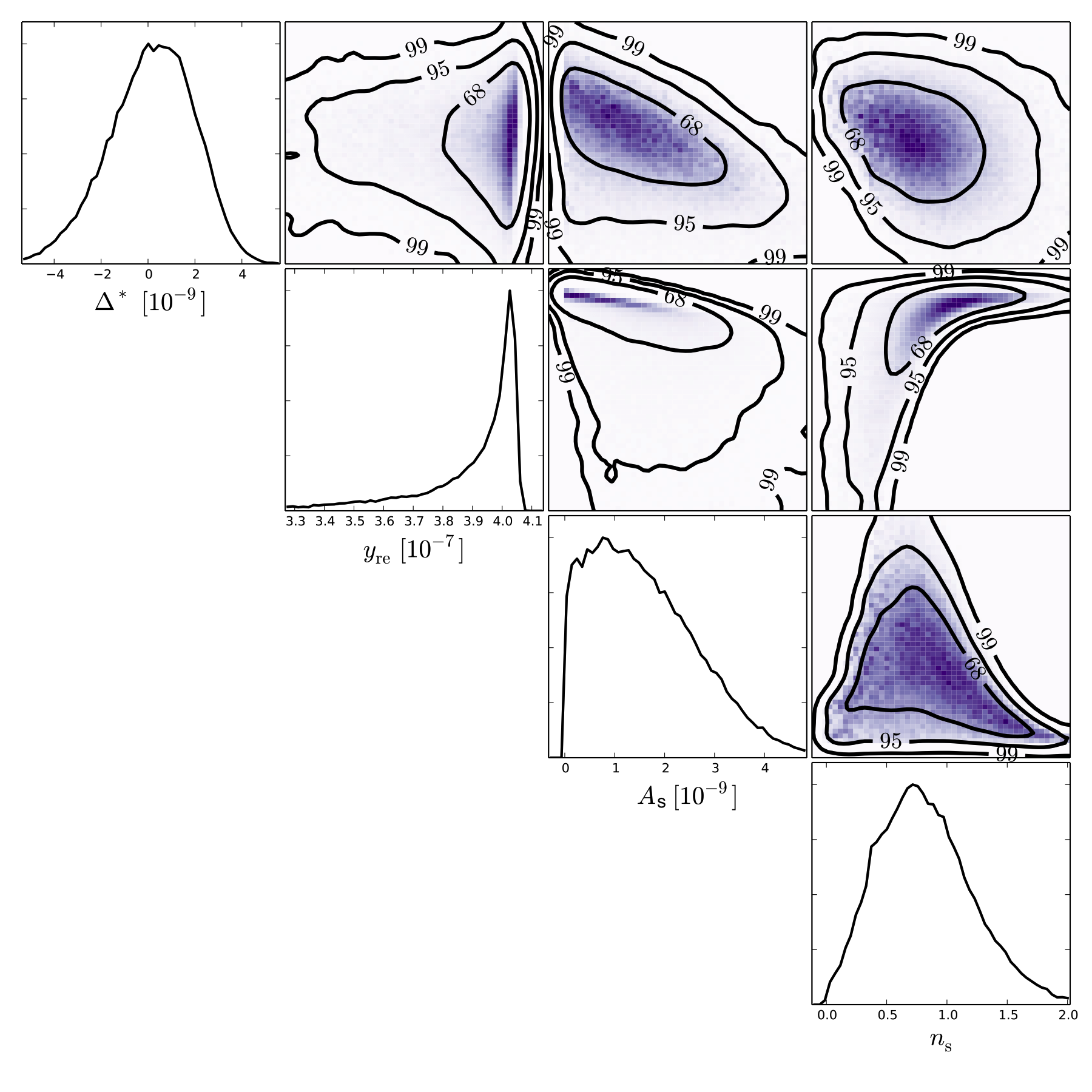}
\caption{ \label{fig:MCMCdist}  Forecast for the parameters $ A_{\rr s}$, $\ns$, $y_{\rr{re}}$ and $\Delta^* \equiv \Delta_T$ at $\kd = 42 \ \Mpc^{-1}$, for PIXIE configuration, using a MCMC sampling method with the \Greens code~\cite{Chluba:2013vsa}. The figure shows 1-D and 2-D posterior marginalized probability density distributions, for a fiducial model given by the best fit of Planck ($\ns^{\rr{fid}} = 0.960$ and no running), $A_{\rr s}^{\rr{fid}}( \kd = 42 \ \Mpc^{-1}) = 1.68 \times 10^{-9}$, $\Delta^{*\rr{fid}} = 0$ and $y^{\rr{fid}}_{\rr{re}} =4 \times 10^{-7} $.}
\end{center}
\end{figure}

Notice however that the PIXIE sensitivity is not sufficient to detect the distortions induced by single field inflation models with no significant running and where the scalar power spectrum on CMB anisotropy scales can be extrapolated down to smaller distortion scales.   Forecasts for the amplitude of the scalar spectrum and its spectral index at the pivot scale $\kd= 42 \Mpc^{-1}$, that lies in the middle of the range of perturbation modes that can be probed with CMB distortions, have been calculated in Ref.~\cite{Khatri:2013dha} with a Fisher matrix method.   Here we have used a bayesian MCMC approach based on the \Greens code~\cite{Chluba:2013vsa} for scalar power spectrum parameter estimation, using an identical pivot scale.  The fiducial values for the scalar power spectrum amplitude and spectral index are given by the best fit of Planck, for a model with no running.  We marginalize also over $y_{\rr {re}}$ corresponding to the $y$-distortions generated during reionization and large scale structure formation, as well as over $\Delta^* \equiv \Delta_T$, where $\Delta_T = (T - T_0) /T_0$ is the variation in the
measurement of the CMB monopole temperature $T_0$, for which we take the fiducial
value $T_0=2.7263 {\rm K}$.
The marginalized posterior probabilities are reported on FIG.~\ref{fig:MCMCdist}.  We find that the shape of the likelihood is strongly non-Gaussian in the plane $(A_{\rr s} - \ns) $.  This behavior induces important deviations compared to the Fisher approach, which highlights the importance of using bayesian analysis for parameter estimation in the context of CMB distortions.  

For spectral index values close to unity around the scale $\kd$, $0.9 < \ns \lesssim 1.1$, we have derived the marginalized posterior probability for the scalar power spectrum amplitude, taking a fiducial value 
$\mathcal P_\zeta^{\rr{fid}} (\kd = 42 \ †\Mpc^{-1}) = 1.68 \times 10^{-9}$, as expected for inflation with no-running and $\ns^{\rr{fid}} = 0.96$.  We find that a scalar power spectrum amplitude 
\be  \label{eq:limit_PIXIE}
\mathcal P_\zeta (\kd = 42 \ \Mpc^{-1}) \approx 2.8 \times 10^{-9} \   \rr{ for \ PIXIE}\,.
\ee
will be disfavored at 95 \% C.L..  We take this value as the lowest scalar power spectrum amplitude that can be distinguished at 95 \% C.L. from the fiducial model with no-running.  

 At this point it must be noticed that the Polarized Radiation Imaging and Spectroscopy Mission (PRISM)~\cite{Andre:2013nfa,Andre:2013afa} has been proposed last year as an L-class ESA mission.  It has been rejected since then but in its best instrumental configuration PRISM is a good exemple of a possible ultimate next to next generation of CMB distortion experiment, with a sensitivity of the order of ten times better than PIXIE
\be
\delta I_\nu ^{\rr{PRISM}} = 6.5 \times 10^{-27} \rr{W m^{-2} Sr^{-1} Hz^{-1}}
\ee 
for frequencies $\nu < 600 \rr{GHz}$ that are the most relevant for the study of CMB distortions.   The $1\sigma$ sensitivity of PRISM is displayed in FIG.~\ref{fig:gendist}.  Such an experiment will allow the measurement of the scalar power spectrum on distortion scales with a good accuracy, improving significantly the current constraints on the running of the spectral index~\cite{Chluba:2013pya}.  Nevertheless, PRISM will not be accurate enough to test the most general single field models where the running values are of second order in slow-roll parameters. 

The main objective of the next two sections is to identify inflationary scenarios leading to an observable level of spectral distortions in the CMB, assuming that the detectable limits are those given by Eq.~(\ref{eq:limit_PIXIE}).   We also comment on the possible detectability by an ultimate experiment like PRISM.  We leave for future work the important questions of foreground removal, how to extract the primordial power spectrum signal from other sources of distortions, as well as the derivation of accurate model specific forecasts using Fisher matrix or Bayesian Markov-Chain-Monte-Carlo (MCMC) methods.  


\section{Testing single-field inflation with CMB distortions}  \label{sec:single_field}

In this Section, we first formulate a series of increasingly narrow
criteria for identifying single-field inflation models that may lead to an observable level of CMB distortions.  Applying these criteria to all the possible regimes of the 49 single-field models listed in Ref.~\cite{Martin:2013tda}, we find that only a few of these may in
principle yield CMB distortions at a level that is observable by the proposed experiments.  For each of these models, we then preform a complete study of the parameter space and determine the regions predicting an increase of power for the curvature perturbations on scales relevant for CMB distortions.  By using a modified version of the \idistort code~\cite{Khatri:2012tw}, the spectrum of $i$-type and $\mu$-type distortions is then calculated for some relevant parameter sets and it is compared to the expected sensitivity of the PIXIE experiment.  We did not consider the $y$-distortions, since these are degenerate with the dominant contribution from the SZ effect and thus very difficult to extract.

\subsection{Necessary conditions for observable CMB distortions }  \label{sec:crit}

The homogeneous dynamics of single field inflation models is governed by the Friedmann-Lema\^itre equations
\be \label{eq:FL1}
H^2 = \frac{1}{3 \Mpl^2} \left[ \frac 1 2 \dot \phi^2 + V(\phi) \right]\,,
\ee 
\be \label{eq:FL2}
\frac{\ddot a}{a} = \frac{1}{3 \Mpl^2} \left[ - \dot \phi^2 + V(\phi)  \right] 
\ee 
and the Klein-Gordon equation 
\be \label{eq:KG}
\ddot \phi + 3 H \dot \phi = - \frac{\dd V}{\dd \phi}~,
\ee
where $\phi$ is the scalar field, $V(\phi)$ its potential, $H$ is the Hubble expansion rate, and $\Mpl$ is the reduced Planck mass.  A dot denotes a derivative with respect to the cosmic time $t$.  For the dynamics during inflation, it is common practice to use the slow-roll approximation, amounting to neglecting the kinetic terms in Eqs.~(\ref{eq:FL1}) and~(\ref{eq:FL2}) as well as the second time derivative of the field $\phi$ in Eq.~(\ref{eq:KG}).  One then defines slow-roll parameters $\epsilon_0 \equiv H_{\rr{ini}} / H$ and $\epsilon_{n\geq 1} \equiv \dd \ln  |\epsilon_{n-1}| / \dd N $, which can be related to the field potential and its derivatives in the slow-roll approximation.  

At first order in slow-roll parameters, the power spectrum amplitude of scalar perturbations and its spectral index are respectively given by~\cite{Stewart:1993bc} 
\begin{align}
\label{powerspectrum}
\mathcal P_\zeta (k) =& \frac{H_k^2}{8 \pi^2 \Mpl^2 \epsilon_{1k}} \left[ 1 - 2 (C + 1) \epsilon_{1k} - C  \epsilon_{2k} \right]\,,
\\
\label{tilt}
\ns(k) =& 1- 2 \epsilon_{1k} - \epsilon_{2k}\,,
\end{align}
with $C \equiv \gamma_{\rr E} + \ln 2 -2 \simeq -0.7296$ and $\gamma_{\rr E}$ being the Euler constant, and where the subscript $k$ indicates that the quantity must be evaluated at the time $t_k$ of Hubble exit of the corresponding mode, i.e. when $k = a(t_k) H(t_k)$.  In the slow-roll approximation, the first and second slow-roll parameters $\epsone$ and $\epstwo$ are related to the field potential and its first and second derivatives,
\begin{align}
\epsone =& \frac{\Mpl^2}{2} \left( \frac{\frac{\dd V}{\dd \phi}}{V}  \right)^2\,,\\
\epstwo =& 2 \Mpl^2 \left[  \left( \frac{\frac{\dd V}{\dd \phi}}{V}  \right)^2 - \frac{1}{V^2}\frac{\dd^2 V}{\dd \phi^2} \right]\,.
\end{align}
The power spectrum amplitude and spectral index are strongly constrained by the recent Planck results~\cite{Ade:2013lta}.  At the pivot scale $\kp = 0.05 \ \Mpc^{-1}$, which exits the Hubble radius about 60 e-folds before the end of inflation, one has
\be
\label{values:fiducial}
\mathcal P_\zeta(\kp) = 2.20 \pm 0.06 \times 10^{-9} , \hspace{10mm} \ns(\kp) = 0.960 \pm 0.007~.
\ee
As already mentioned, the main interest in searching for CMB distortions from inflation results from the fact that these probe a different range of modes than accessible from CMB anisotropies. However, PIXIE is expected to detect only spectral distortions from curvature perturbations with an amplitude that is larger on smaller scales than on the scales probed by CMB anisotropies.  Considering this restriction, one can impose three necessary criteria for an inflation model to i) satisfy the Planck constraints, and ii) to lead to an observable level of CMB distortions:
\begin{enumerate}
\item There exists a phase where  $\epstwo < -2 \epsone < 0 $ : With this condition we identify models for which at some point during the scalar field evolution there is an increase of power in the spectrum of curvature perturbations (i.e. a locally blue spectrum).  Imposing this restriction already eliminates large classes of models like large field inflation, Higgs inflation, natural inflation, exponential SUSY inflation, Logamediate inflation (for which $\epstwo <0$ is possible but with $\epstwo \gtrsim - 2 \epsone$ only).
\item A phase with $\ns < 1$ must be followed by a phase with $\ns > 1$ : With this condition, we specify that the increase of power (blue spectrum) must occur after a phase where the spectral index is in accordance with CMB anisotropy observations (i.e. the spectrum is red
at the scale $\kp$).  One therefore eliminate models where $\ns > 1$ at all times, and models where the phase corresponding to $\ns >1$ takes place only at times earlier than the time of Hubble exit of the pivot scale $\kp$ (e.g. Brane SUSY breaking inflation, Supergravity Brane inflation).   An additional important criterion is that
$\epsone \neq  0$ (in other words, there are no points where the slope of the potential is vanishing) between these two phases, because otherwise an infinite (or extremely large if one considers also the quantum stochastic field evolution) number of e-folds separates these two phases.  This additional criterion permits to exclude Inflection Point inflation and MSSM Inflation models.  
\item $\ns = 0.960 \pm 0.007 $ at $\kp = 0.05 \Mpc^{-1}$ and $\ns >1$ at $\kd= 42 \Mpc^{-1}$:  The first condition is required for the model to satisfy the Planck constraints whereas the second one is a sufficient, but not necessary, condition for an increase of power on scales relevant for CMB distortions.  As already mentioned, the scale $\kd= 42 \Mpc^{-1}$ lies in the middle of the range of perturbation modes that can be probed with CMB distortions.   Our choice to $\kd$ is arbitrary and a sufficient enhancement of power on distortion scales compared to the power on CMB anisotropy scales can in principle be obtained if $\ns > 1$ on scales $\kp < k < \kd$.  In practice however, we did not find any single-field model where the enhancement of the scalar power spectrum occurs only for a restricted number of scales within this range.  We do not impose a condition on the power spectrum amplitude at the pivot scale, which can be accommodated by a simple rescaling of the potential without modifying the field evolution $\phi(N)$, where $N$ ia the number of e-folds.    
\end{enumerate}

These three criteria are increasingly narrow, such that they entail an effective procedure for selecting those models that may lead to distortions detectable by PIXIE.
The third criterion is the most restrictive but also the most difficult to apply.  It requires the integration of field trajectories in order to evaluate the scalar power spectrum and the spectral index values on the relevant scales.  Nevertheless, already by imposing the first two criteria, we exclude most single field models and a short list of only five models remains: Hybrid inflation in the valley (VHI), the generalised Minimal Supersymmetric model (GMSSM), the generalised renormalisable inflection point model (GRIP) and the running mass (RM) and non-canonical K\"ahler inflation (NCKI) models.  The parameter space of these models is explored in detail in the following Sections, observable predictions are derived and regions leading to potentially observable CMB distortions are determined. 

\subsection{Hybrid inflation in the Valley (VHI)}

\begin{figure}
\begin{center}
\includegraphics[width=8.5cm]{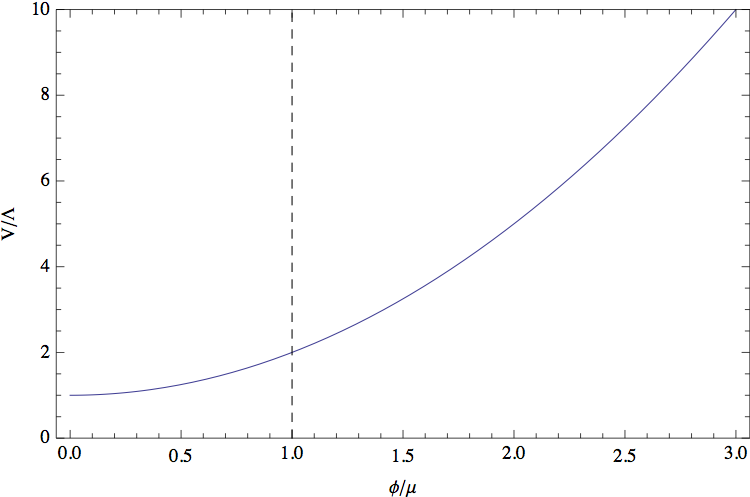}~~~~\includegraphics[width=8.5cm]{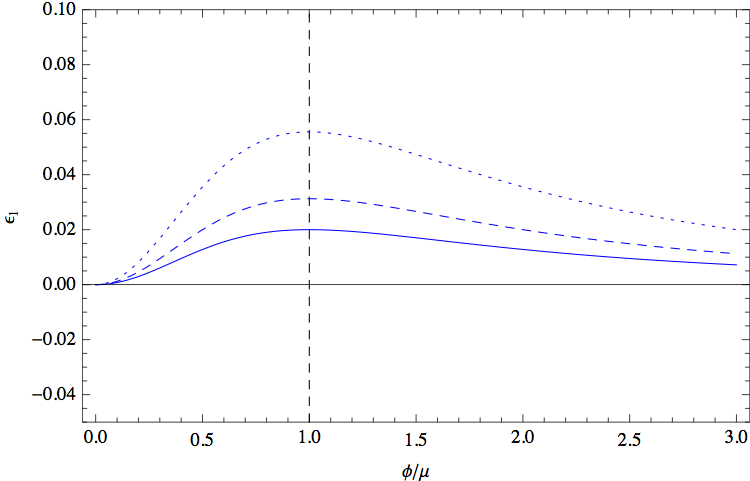}
\includegraphics[width=8.5cm]{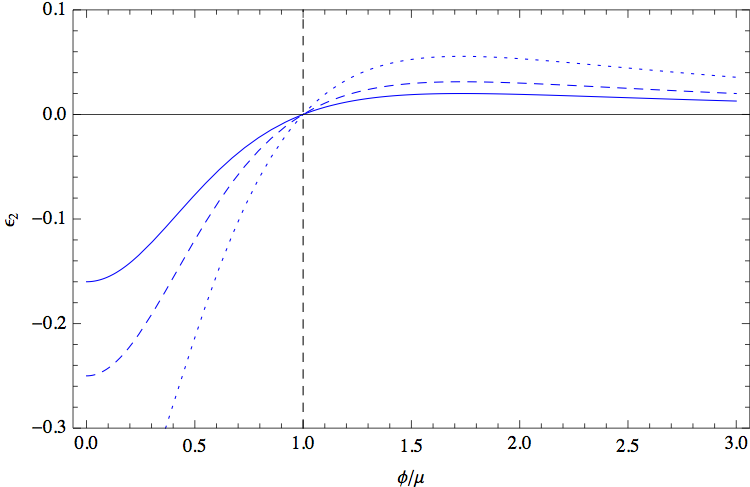}~~~~\includegraphics[width=8.5cm]{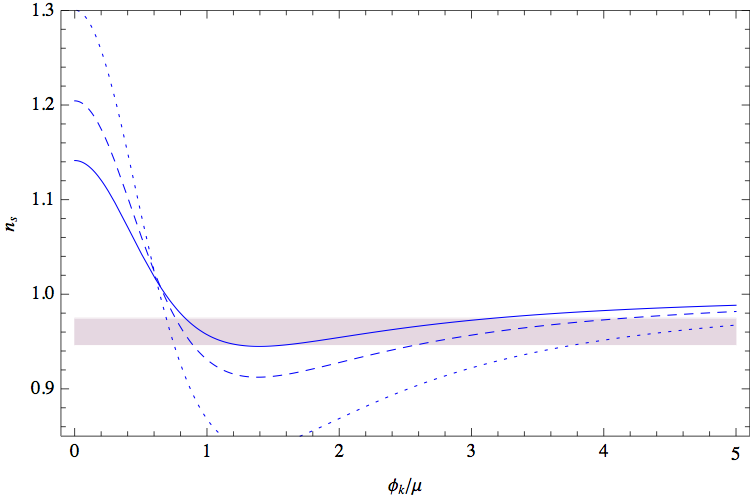}

\caption{\label{fig:pot_HIV}
Top left: original hybrid potential.  Top right and bottom left:  first and second 
slow-roll parameters $\epsilon_1$ and $\epsilon_2$.
Bottom right: scalar spectral index $\ns$ as a function of $\phi_k$. The horizontal band corresponds to the Planck $95 \%$ C.L. constraints.
The parameter values in the diagrams for $\epsilon_{1,2}$ and $n_s$ are $\mu_{\rr{VHI}} = 3 \Mpl$ (dotted), $\mu_{\rr{VHI}} = 4 \Mpl$ (dashed) and $\mu_{\rr{VHI}} = 5 \Mpl$ (solid). We are interested in the region close to $\phi_k /\mu_{\rr{VHI}} = 1$, where the power spectrum of curvature perturbations is red-tilted, but which becomes blue-tilted a few e-folds later (i.e. at smaller values of $\phi_k$.) } 
\end{center}
\end{figure}

The original hybrid potential~\cite{Linde:1993cn} has a nearly flat direction
with non-vanishing potential energy. Along this direction, there is a critical point
that separates a valley from a region of tachyonic instability.
Inflation occurs within the valley until the inflaton crosses the critical point,
which triggers the so-called waterfall
phase during which the scalar-field configuration  settles at the global minimum
of the potential. In the common picture, the waterfall phase is nearly instantaneous (contrary to the mild waterfall case studied in Sec.~\ref{sec:hybrid2field})  and the dynamics along the valley is described by the effective single-field potential
\be
\label{V:VHI}
V(\phi) = \Lambda \left( 1+ \frac{\phi^2}{\mu_{\rr{VHI}}^2} \right)\,,
\ee
which is of a very simple form that we illustrate in Fig.~\ref{fig:pot_HIV}.
Notice that the inflationary valley can be reached
by field trajectories exterior to it without specific fine-tuning of initial conditions~\cite{Clesse:2008pf,Clesse:2009ur,Clesse:2009zd,Easther:2013bga}.   One of the attractive features of the hybrid model is that many e-folds of expansion (much more than 60) can be realized for sub-Planckian field values, such that possible corrections
from gravitational interactions can be ignored in the effective field theory description. This regime corresponds to $\mu_{\rr{VHI}} \gg \Mpl$, but note however that the inflaton mass $m^2 = 2 \Lambda / \mu_{\rr{VHI}}^2$ takes values lower than the Planck mass when imposing the observed normalisation of
the power spectrum.  Depending on the dominant contribution to
the potential, we refer to the small field regime as vacuum dominated
and to the large field regime as field dominated. In the case where $\mu_{\rr{VHI}} \lesssim 1.5 \Mpl$, the slow-roll conditions are violated at the transition between the large field ($\phi > \mu_{\rr{VHI}}$) and small field phases ($\phi < \mu_{\rr{VHI}}$), such that the inflaton field acquires a sufficient amount of kinetic energy to prevent inflation from taking place in the small field regime~\cite{Clesse:2008pf}. In this situation, all of the last 60 e-folds of inflation occur at large field values, and the spectral index is red and possibly in agreement with CMB observations. 

In order to avoid the regime where the slow-roll dynamics is violated at the transition between
the large and small field phases, we will focus on parameter values $\mu_{\rr{VHI}} \gtrsim 1.5 \Mpl$.   
The slow-roll parameters are given by
\begin{subequations}
\label{eps:HIV}
\begin{align}
\epsilon_1 =& \frac{2 \Mpl^2 x^2 }{\mu^2 \left( 1 + x^2 \right)^2}~,\\
\epsilon_2 =&  \frac{4 \Mpl^2 \left( -1 + x^2 \right) }{\mu^2 \left( 1 + x^2 \right)^2}~,
\end{align}
\end{subequations}
where we have defined $x \equiv \phi / \mu_{\rr{VHI}}$.    They are plotted in  Fig.~\ref{fig:pot_HIV} for various values of the parameter $\mu_{\rr{VHI}}$.  
In the vacuum dominated regime, since $\epsilon_1 \ll 1$ and $\epsilon_2 < 0$, a blue spectrum of scalar perturbation is expected.   Inflation stops at the critical point $\phi_c$ (which is thus a third model parameter) below which the potential develops the tachyonic instability.

\begin{figure}
\begin{center}
\includegraphics[width=12.cm]{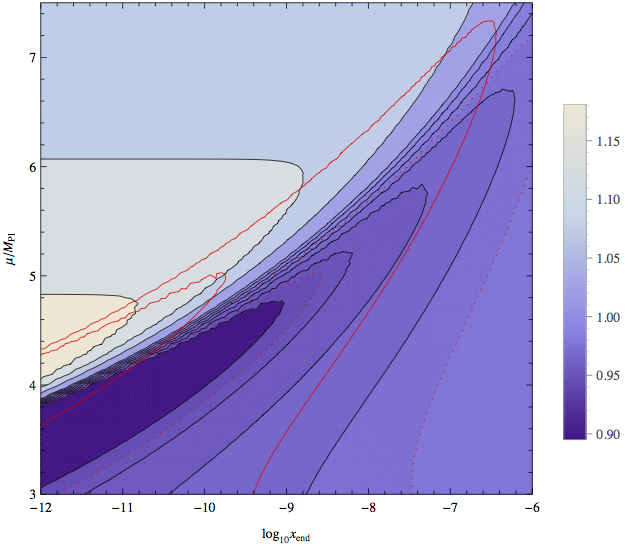}
\caption{\label{fig:nsplane2_HIV}Contours of spectral index values from $n_s = 0.94$ up to $n_s = 1.15$ for the original hybrid model with $N_{\kp} = 60$ in the plane $\mu_{\rr{VHI}}$ - $\log_{10} (\phic / \mu_{\rr{VHI}})$, evaluated  at the scale $\kd = 42 \Mpc^{-1}$.   The area
between the solid red contours is consistent with the Planck measurement
of the spectral index at $\kp= 0.05 \ \Mpc^{-1}$ at $95 \%$ C.L..  The corresponding contours evaluated at the scale $\kd$ are included (dotted red) in order to visualize how the spectral index changes when the scale varies.
} 
\end{center}
\end{figure}

\begin{figure}
\begin{center}
\includegraphics[width=12cm]{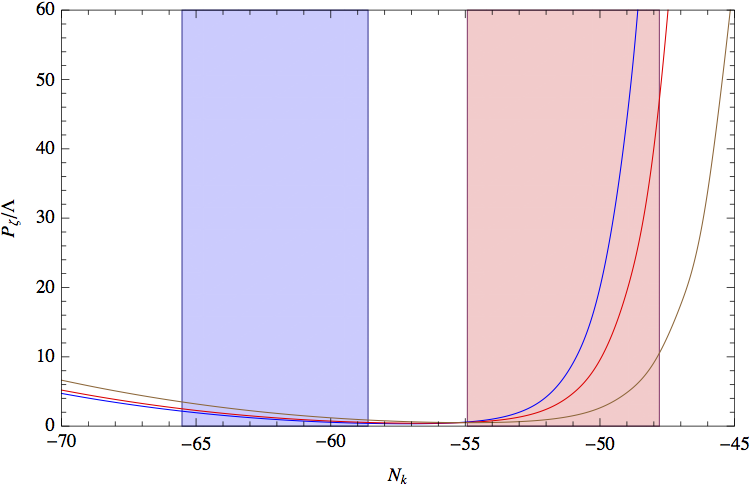}
\caption{\label{fig:Pzeta_HIV}Power spectrum of curvature perturbations for three parameter sets: $\mu_{\rr{VHI}} = 5 \Mpl $ and $\log_{10} (\phic/\mu_{\rr{VHI}}) = -10$ (blue),  $\mu_{\rr{VHI}} = 5.5 \Mpl$ and $\log_{10} (\phic /\mu_{\rr{VHI}}) = -9$ (red),  $\mu_{\rr{VHI}} = 6. \Mpl$ and $\log_{10} (\phic /\mu_{\rr{VHI}}) = -8 $ (brown).  The blue region corresponds to scales observable with standard CMB observations whereas the red region corresponds to scale that can be probed with CMB distortions.
The amplitude can be normalised to $\mathcal P_\zeta (\kp) = 2. \times 10^{-9}$ by a suitable choice of the parameter $\Lambda$. 
}
\end{center}
\end{figure}

\begin{figure}
\begin{center}
\includegraphics[width=12cm]{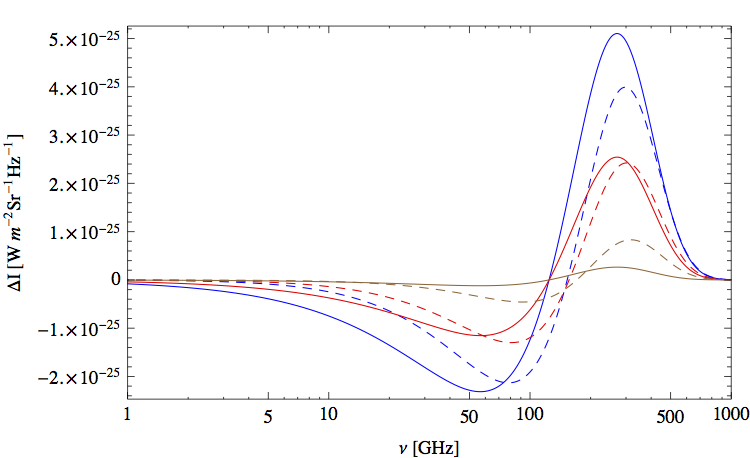}
\caption{\label{fig:distord_HIV}Intermediate (dashed) and $\mu$-type (solid) spectral distortions as a function of the frequency, for the original hybrid model and the parameter sets as in Fig.~\ref{fig:Pzeta_HIV}.  They correspond, respectively, to $\mu = 1.7 \times 10^{-7} $ (blue), $\mu = 8.5 \times 10^{-8}$ (red) and $\mu = 2.5 \times 10^{-8}$ (brown).    The signal is detectable by PIXIE at more than $5\sigma$ in the first two cases, and at about $2 \sigma$ in the latter case.  
}
\end{center}
\end{figure}

Since we look for a model predicting a red spectrum on large scales and a blue spectrum on small scales, we focus on the intermediate regime, for which i) the slow-roll approximation is valid for all field values above the critical instability point $\phi_c$, including trans-Planckian configurations, i.e. we require $\mu_{\rr{VHI}} > 1.5 \Mpl$, ii) scales relevant for CMB anisotropies leave the Hubble radius close to the point where $\phi = \mu_{\rr{VHI}}$, at which $\epsilon_1 = \mathcal O (0.1)$ and $| \epsilon_2 | < \epsilon_1$, such that the spectral index is red and
in agreement with observations, and iii) a few e-folds later, the vacuum dominated regime has just taken over and the spectral index for corresponding modes is blue. An enhancement of power therefore occurs on small scales, which may lead to CMB distortions at an observable level. 

As long as the slow-roll approximation is valid, one can integrate 
\be
\label{Klein:Gordon}
\frac{\dd \phi}{\dd N} = - \Mpl^2\frac{\dd \ln V}{\dd \phi} 
\ee
in order
to get the number of e-folds realized between some field value $\phi$ and the end of inflation at $ \phi_{\rr{end}} = \phic$,
\be
N(\phi) - \Nend = -\frac{\mu_{\rr{VHI}}^2}{4\Mpl^2} \left[x^2+2 \ln (x) \right] + \frac{\mu_{\rr{VHI}}^2}{4\Mpl^2} \left[\xend^2+2 \ln (\xend) \right]\,.
\ee
This relation can be inverted to get the field value $\phi_k$ at which the perturbation mode $k$ leaves the Hubble radius, $N_k$ e-folds before the end of inflation.  For simplicity we have set $N_{k_{\rr{p}}} = 60$, independent of the energy scale of inflation and the details of the reheating history.  Once $\phi_k$ is obtained, the power spectrum amplitude and the spectral index for the mode $k$ can be easily calculated by using Eqs.~(\ref{powerspectrum}), (\ref{tilt}) and (\ref{eps:HIV}).

In order to decide whether CMB distortions at an observable level may be produced,
we now apply the third criterion from Section~\ref{sec:crit}. For this purpose,
we evaluate the spectral index in the plane ($\mu_{\rr{VHI}}$, $x_{\rr{end}}$) for the two pivot scales $\kp = 0.05 \ \Mpc^{-1}$ and $\kd= 42 \Mpc^{-1}$. The latter scale lies within the range of modes that can be probed with CMB distortions, corresponding to $N_{\kd} = N_{\kp} - 6.7$.
In FIG.~\ref{fig:nsplane2_HIV}, we show the contours for the spectral index evaluated at the
scale $\kd$ and superimpose these with
the 95\% C.L. constraints from Planck on the spectral index at the scale $\kp$.   In order to illustrate how the spectral index changes when the scale varies, we have also plotted these contours at the scale $\kd$.  

We find a region of the parameter space that is in agreement with Planck data and leads at the same time
to an increase of power on CMB distortion scales. As it can be seen from FIG.~\ref{fig:nsplane2_HIV}, this
region corresponds to a band going from $\mu_{\rr{VHI}} \approx 4.2 \Mpl$ and $\log_{10} \xend \approx -12$ up to  $\mu_{\rr{VHI}} \approx 7 \Mpl$ and $\log_{10} \xend \approx -7$.   Therefore, the corresponding values of the critical instability point $\phic=\xend \mu_{\rr{VHI}}$ are much below the Planck mass.  

From the region identified above, we select
three representative parameter sets that correspond to increasing values of $n_s (\kd)$.
In FIG.~\ref{fig:Pzeta_HIV}, we plot the resulting scalar power spectrum as a function of
the scale (expressed in terms of e-folds), and in FIG.~\ref{fig:distord_HIV}, we
show the resulting spectrum of distortions. From FIG.~\ref{fig:Pzeta_HIV}, we explicitly
see the enhancement of the scalar power spectrum on distortion scales,  which can be more than one order of magnitude larger than on anisotropy scales.  Given the bound of Eq.~(\ref{eq:limit_PIXIE}) for PIXIE, parameters leading to a maximal enhancement are thus detectable at more than $5\sigma$ confidence level.   This is translated into spectral distortions shown in FIG.~\ref{fig:distord_HIV}, up to one order of magnitude larger than those expected for a model with constant $\ns \simeq 0.96$.    The resulting $\mu$-distortions correspond to $\mu \gtrsim 10^{-7}$ in the case where the enhancement is maximal, which is as expected far above the $5\sigma$ detection level of PIXIE.   The parameter space leading to a clear detection of distortions is located along the band identified above, with a critical instability point located at $ \phi_c \lesssim 10^{- 9}†\Mpl$.   One could nevertheless argue that those $\mu$ distortions cannot be distinguished from possible other sources, such as dark matter decay/annihilation.   For this reason we have also plotted the spectrum of intermediate-type distortions, from which it should be possible to distinguish between different origins of the signal.  We find that in the case of a limited enhancement of the spectrum the intermediate-type distortions dominate over the $\mu$-type.   For PIXIE, whose sensitivity is given in Eq.~(\ref{sensitivity:PIXIE}), the signal is detectable at more than $5\sigma$ along the band with $\phic \lesssim 10^{-9} \Mpl$, and at about $2 \sigma$ if $\phic \lesssim 10^{-8} \Mpl$.   In the case of a strong enhancement the $\mu$-distortion signal is dominant, but the level of intermediate distortions does not decrease and therefore remains clearly detectable.  The smallest scales re-enter inside the horizon earlier than larger scales and therefore contribute to energy injection in the primeval plasma at earlier times.  Enhancing the power spectrum on these scales therefore corresponds to more $\mu$-type distortions, whereas enhanced power on the largest distortion scales induce more intermediate-type distortions.   In the case of the hybrid regime, depending on the parameters, the two processes compete.  But as illustrated on Fig. 5, the increase of power is close to exponential.  This implies that when the distortion signal is maximal, the major contribution comes from the smallest scales and it is dominated by the $\mu$-type distortions, contrary to the simplest model of constant spectral index for which the signal is dominated by intermediate-type distortions.
 With a ten times better sensitivity, a PRISM-class experiment would not only detect the distortion signal but should also allow precise parameter estimation and would extend the probed parameter space to the entirety of the band identified above, corresponding to values of $\phic$ up to the $\phic \sim 10^{-6} \Mpl$. 

We therefore conclude that the original hybrid model in the fast waterfall regime can lead to CMB distortions observable by PIXIE, but only if the inflaton field takes super-Planckian
expectation values and only in a rather tuned region of the parameter space, imposed by the
condition that the transition between the field-dominated and the vacuum dominated regime
occurs around $N_k\approx 55$ e-folds before the end of inflation.

\subsection{Non-canonical K\"ahler Inflation (NCKI)}

\begin{figure}
\begin{center}
\includegraphics[width=8.5cm]{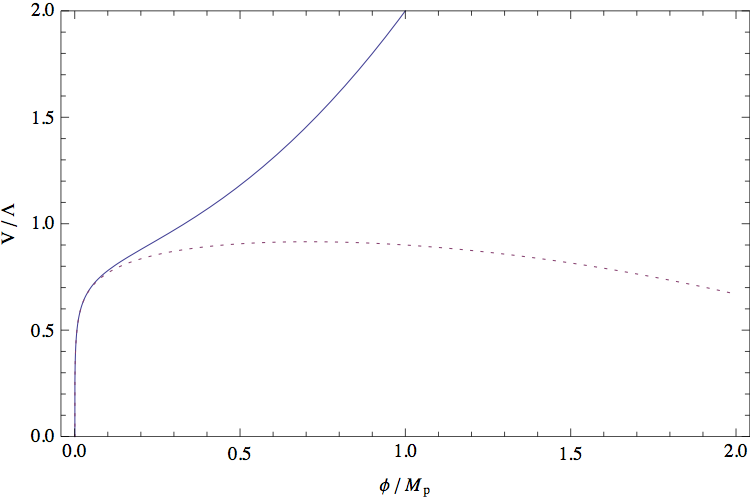}~~~~\includegraphics[width=8.5cm]{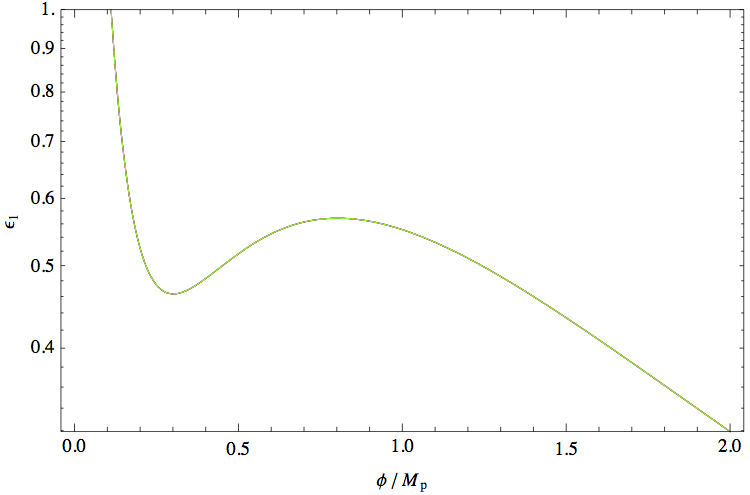}
\includegraphics[width=8.5cm]{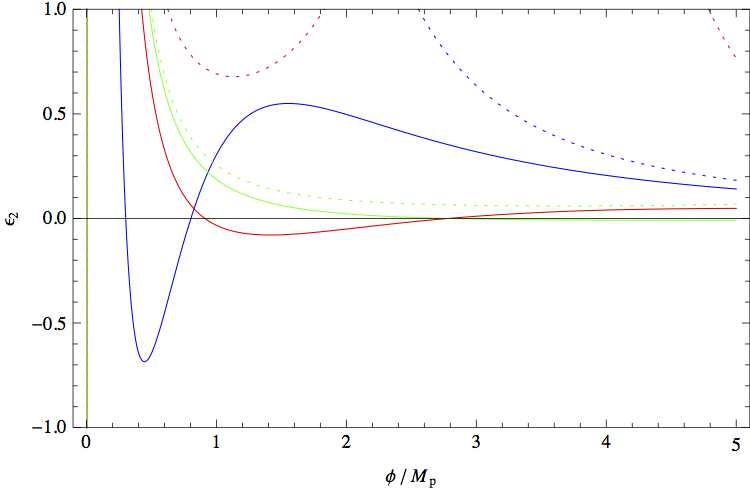}~~~~\includegraphics[width=8.5cm]{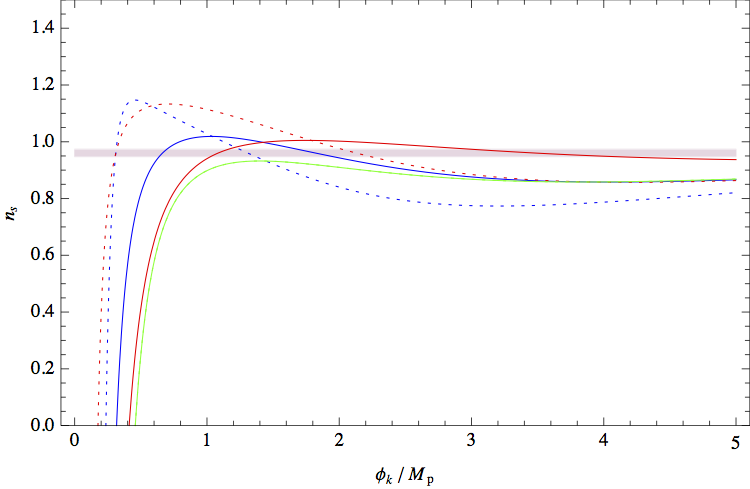}
\caption{\label{fig:pot_NCKI}Top left: potential of the NCKI model for the
parameters $\alpha=0.1$, $\beta=1$ (blue, solid) and $\alpha=0.1$, $\beta=-0.1$ (red, dotted).
Top right and bottom left: first and second slow-roll parameters $\epsone,\epstwo$.
 Bottom right: spectral index value as a function of $\phi_k$. The parameters in the
 plots for $\epsone,\epstwo,\ns$ are: $\alpha=0.1$, $\beta=1$ (blue, solid);  $\alpha=0.1$, $\beta=-1$ (blue, dotted); $\alpha=0.1$, $\beta=0.1$ (red, solid); $\alpha=0.1$, $\beta=-0.1$ (red, dotted); $\alpha=0.1$, $\beta=0.01$ (green, solid); $\alpha=0.1$, $\beta=-0.01$ (green, dotted).} 
\end{center}
\end{figure}

\begin{figure}
\begin{center}
\includegraphics[width=12.cm]{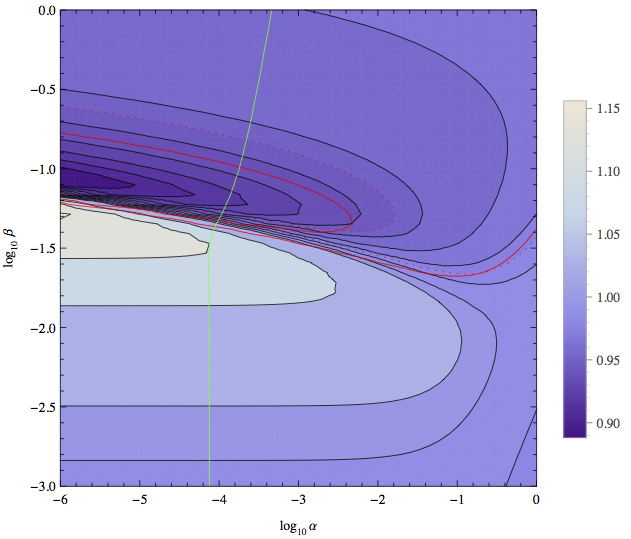}
\caption{ \label{fig:ns_NCKI}Spectral index value in the parameter space $(\alpha, \beta)$ of the NCKI model, at $\kd = 42\ \Mpc^{-1}$.  Between the
red contours is the area allowed by the Planck constraints on $\ns$ for the pivot scale of CMB anisotropies $\kp = 0.05 \ \Mpc^{-1}$. The same contours for the scale $\kd$ are added (dotted red) to visualize how the spectral index changes when the scale varies.  To the left of the green, solid and almost vertical line, the approximation
of the radiative correction to the potential in the form~(\ref{V:NCKI}) is not valid.}
\end{center}
\end{figure}

\begin{figure}
\begin{center}
\includegraphics[width=8.5cm]{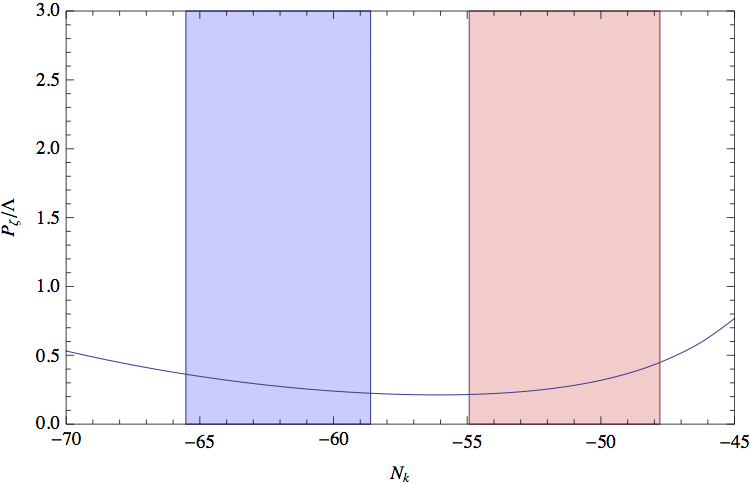}
\caption{\label{fig:Pzeta_NCKI}Power spectrum of curvature perturbations for the NCKI model with $\alpha = 10^ {-6}$ and $\beta = 10^ {-1.2}$.  Those parameter values lie in the region where the approximation of the radiative corrections is not valid but are representative of the strongest possible enhancement of power on CMB distortion scales considering an effective potential of the form of Eq.~\ref{V:NCKI}.  }
\end{center}
\end{figure}

\begin{figure}
\begin{center}
\includegraphics[width=9.5cm]{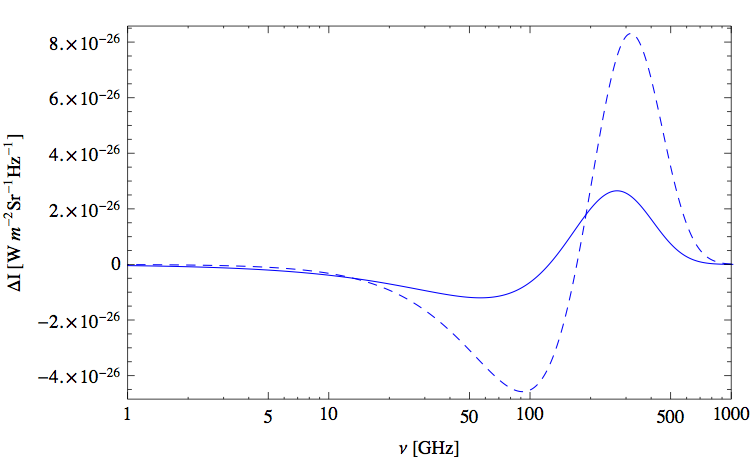}
\caption{\label{fig:distort_NCKI}Spectrum of intermediate and $\mu$-distortions (respectively dashed and solid line) for the NCKI model with $\alpha = 10^ {-6}$ and $\beta = 10^ {-1.2}$.  They correspond to $\mu = 8.9 \times 10^{-9}$ and $y = 5.5 \times 10^{-9}$ and are below the sensitivity of PIXIE.   They are nevertheless within the range of detectability of a PRISM-class experiment.} 
\end{center}
\end{figure}

In the usual F-term and D-term hybrid models~\cite{Dvali:1994ms,Binetruy:1996xj,Halyo:1996pp,Garbrecht:2006az,Clauwens:2007wc,Kallosh:2003ux}, the flat direction of the potential is lifted by radiative loop corrections as well as supergravity corrections induced by a minimal (also referred
to as canonical)
K\"ahler potential. Additional corrections to the potential may arise from higher order operators.    If the K\"ahler potential is non-canonical, there occurs an extra mass term for the
inflaton, such that the potential can become hill-top like.  It reads
\be
\label{V:NCKI}
V(\phi) = \Lambda \left( 1 + \alpha \ln x + \beta x^2 \right)\,,
\ee
where $x = \phi / \Mpl$. The dimensionless parameter $\alpha$ is positive, whereas $\beta$ is
expected to be of $\mathcal O(1)$ and can either be positive or negative.  Setting $\beta >0$, the potential increases monotonously with $|\phi|$.  Field trajectories can slow-roll back along the potential and inflation ends with a waterfall instability.  Setting $\beta < 0$, the potential exhibits a maximum and from this point the trajectoris can evolve either in the direction of smaller or in the direction of larger field values.   The first and second slow-roll parameters are given by~\cite{Martin:2013tda}
\begin{subequations}
\begin{align}
\epsone =& \frac{\left(\alpha +2 \beta  x^2\right)^2}{2 \left[ \alpha  x \log (x)+\beta  x^3+x\right]^2}\,,\\
\epstwo =& \frac{2 \left[  (5 \alpha -2) \beta  x^2+\alpha  \log (x) \left(\alpha -2 \beta  x^2\right)+\alpha  (\alpha +1)+2 \beta ^2 x^4\right]}{\left[\alpha
    x \log (x)+\beta  x^3+x\right]^2}
\end{align}
\end{subequations}
and are represented in FIG.~\ref{fig:pot_NCKI} for several values of the parameters.  Evidently, the second slow-roll parameter $\epstwo $ can take negative values only if $\beta >0$.  By applying the first criterion given in Sec.~\ref{sec:crit}, we can thus restrict to the case of positive $\beta $. In the limit $\alpha \rightarrow 0$, one recovers the original hybrid potential, whereas in the limit $\beta \rightarrow 0$ one recovers the potential with
a purely radiatively lifted flat direction. The NCKI potential therefore encompasses a rather large variety of models, in particular supersymmetric F-term inflation.  The spectral index, which is  represented in FIG.~\ref{fig:pot_NCKI}, takes values lower than unity at large fields and close $\phi =0$.  For modes exiting the Hubble radius during the transition between these two regimes, the power spectrum is blue. There exist two ways for inflation to stop in the NCKI model:  i) the field trajectory reaches the critical point $\phi_c$ where the potential develops the waterfall instability, which we will assume to be instantaneous, ii)  the slow-roll conditions are violated. For the second case, we take $\epstwo = 1$  as the condition
to identify when the end of inflation occurs. This is well approximated by the time
when $\phi$ reaches the value
\be
\phi_{\rm end} = \sqrt{2}\alpha \,.
\ee

Integrating the Klein-Gordon equation~(\ref{Klein:Gordon}) for the scalar field in the slow-roll approximation, we numerically obtain the number of e-folds and then the field values at Hubble exit of the pivot scale.  The predictions for the spectral index value in the 2-dimensional parameter space, for the distortion scale $\kd = 42 \Mpc^{-1}$ is presented in FIG.~\ref{fig:ns_NCKI}. Again, we superimpose this with the $95\%$ C.L. Planck constraints on the spectral tilt at the pivot scale $\kp=0.05 \Mpc^{-1}$. We can thus identify the region in parameter space that is in agreement with observations and leads to an increase of power on scales relevant for CMB distortions.  This region corresponds to a very thin band covering several order of magnitudes for the parameter $\alpha$, up to $\alpha \simeq 10^{-3}$, whereas the value of $\beta$ remains in the range $10^{-2} \lesssim \beta \lesssim 10^{-1}$.  
However, we find that the increase of power cannot exceed one order of magnitude [cf. FIG.~\ref{fig:Pzeta_NCKI} and Eqs.~(\ref{eq:limit_PIXIE}) and ~(\ref{fig:distort_NCKI})].

We should also comment that the change from a red to a blue spectral index between the
scales $\kp$ and $\kp$ requires trans-Planckian values of the inflaton field
during the horizon exit of these scales, which might contradict the standard assumptions
about the validity of supersymmetry as an effective field theory, that is only
applicable for field values below the Planck scale. Moreover, the potential~(\ref{V:NCKI})
approximates the radiative corrections in a way that is only applicable
if $\alpha\gg\Lambda^{1/3}/[2^6 \pi^{14/3}\Mpl^{4/3}]$, cf. the detailed form
of the F-term potential in Ref.~\cite{Dvali:1994ms}. We have indicated this constraint
in FIG.~\ref{fig:ns_NCKI}, which would rule out parts the parameter space that is consistent with
the third condition from Sec.~\ref{sec:crit} from the outset.

\subsection{Generalized MSSM Inflation (GMSSM)}

\begin{figure}
\begin{center}
\includegraphics[width=8.5cm]{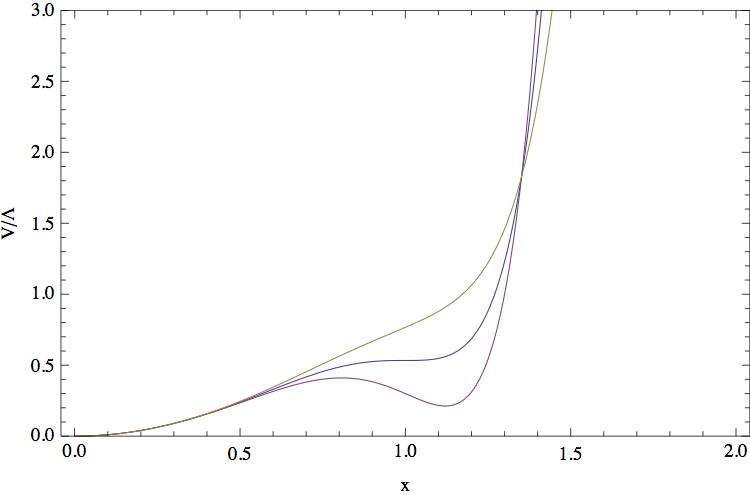}~~~~\includegraphics[width=8.5cm]{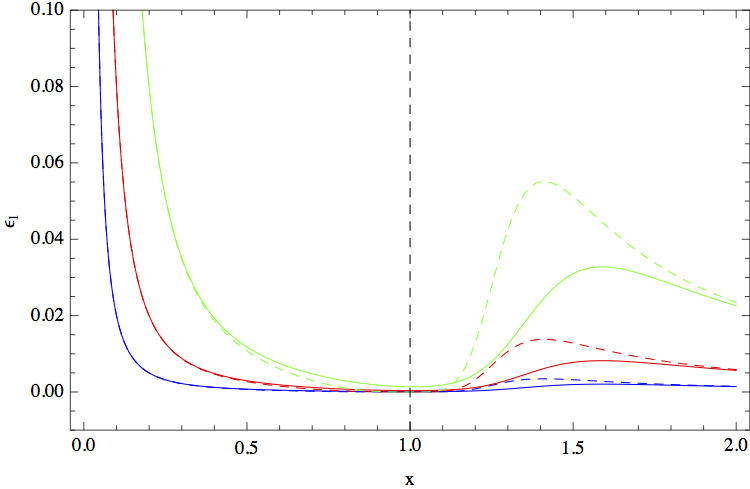}
\includegraphics[width=8.5cm]{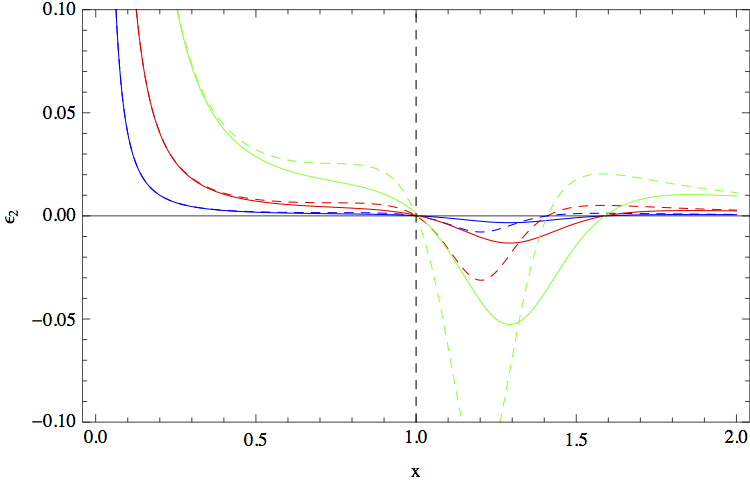}~~~~\includegraphics[width=8.5cm]{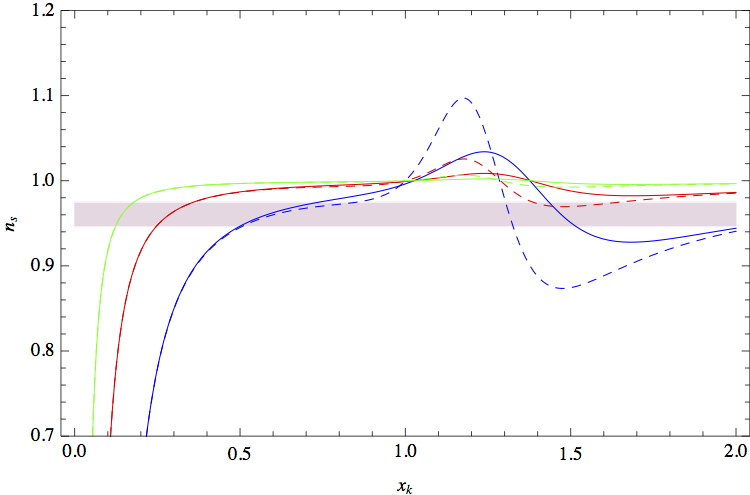}
\caption{\label{fig:pot_GMSSM}Field potential (top left) with $\alpha = 1.5$ (red), $\alpha = 1 $ (blue) and $\alpha = 0.5$ (yellow) for the Generalized MSSM model.  First and second slow-roll parameters (respectively top right and bottom left) and spectral index value as a function of $\phi_k$ (bottom right), for $\phi_0 = (100,50,25) \Mpl$ (respectively blue, red and green) and $\alpha = 0.5 ,1 $ (respectively dashed and solid).  A phase with $\ns > 1$ follows a phase with $\ns < 1$ in the case $\alpha < 1$ and $\phi_0 \gg \Mpl$.}
\end{center}
\end{figure}

\begin{figure}
\begin{center}
\includegraphics[width=12.cm]{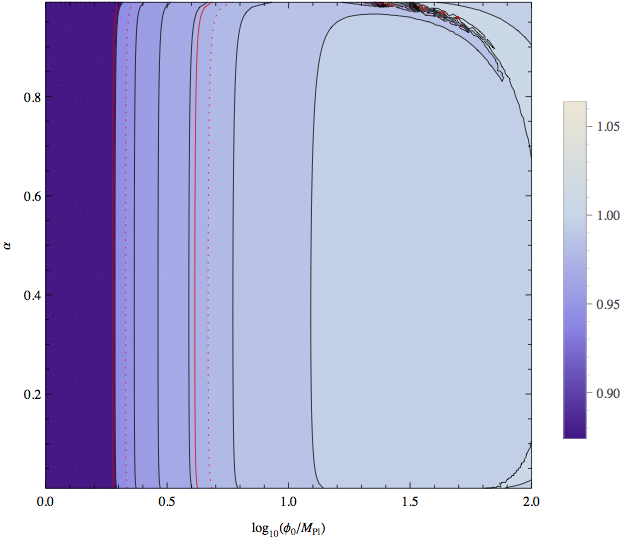}
\caption{\label{fig:ns_GMSSM}Spectral index value in the parameter space for the GMSSM model, for the pivot scale of CMB anisotropies $\kp = 0.05 \ \Mpc^{-1}$ and of CMB distortions $\kd = 42 \ \Mpc^{-1}$.
The area between the solid red contours is in agreement with the Planck constraints.  The same contours for the scale $\kd$ (dashed red) illustrate how the spectral index changes when the scale varies. 
}
\end{center}
\end{figure}

In Ref.~\cite{Allahverdi:2006iq}, it has been proposed that inflation can proceed along a
flat direction of the Minimal Supersymmetric Standard Model (MSSM). These directions are
lifted by soft SUSY breaking terms as well as by non-renormalizable operators that are
expected within a supergravity framework.
In the original model, directions with an exact inflection point, where
the first slow-roll parameter $\epsone$ vanishes, were discussed. This implies that
in the classical slow-roll approximation, an infinite number of e-folds separates the inflationary
phases at field values above and below the inflection point.
As a consequence, we cannot apply the second criterion from Sec.~\ref{sec:crit}. This immediately
gets us to study the Generalized MSSM Inflation (GMSSM)~\cite{Martin:2013tda} scenario,
where the potential only features an approximate inflection point~\cite{Lyth:2006ec}.
Following Ref.~\cite{Martin:2013tda}, we parametrise the GMSSM model as
\be
V(\phi) = \Lambda \left( x^ 2 - \frac 2 3 \alpha x^ 6 + \frac 1 5 \alpha x^{10}   \right)\,,
\ee
where $x \equiv \phi / \phi_0 $ and $\alpha$ is a dimensionless parameter. 
The powers of $x$ are selected by identifying the lowest order soft and non-renormalizable 
operators that are available for lifting the flat direction in the
MSSM~\cite{Allahverdi:2006iq}.

Setting $\alpha = 1$, the original MSSM model with an exact
inflection point is recovered.  The potential is similar to the one of the
GRIPI model below, which differs in the particular powers of $x$.
When $\alpha > 1$, it develops a maximum with three possible inflationary regimes (from the maximum towards decreasing field values, from the maximum towards increasing field values, and from the large field regime in the direction of decreasing field values), whereas for $\alpha < 1$ the potential is monotonous and there are no local extrema
or inflection points. These different possibilities are illustrated in FIG.~\ref{fig:pot_GMSSM}. The slow-roll parameters for the present model read~\cite{Martin:2013tda}
\begin{subequations}
\begin{align}
\epsone =& \Mpl^2\frac{450 \left[\alpha  \left(x^4-2\right) x^4+1\right]^2}{\phi_0^2 x^2 \left[\alpha  \left(3 x^4-10\right) x^4+15\right]^2}\,,
\\
\epstwo =& \Mpl^2 \frac{60 \left(\alpha  x^4 \left\{ x^4 \left[ \alpha  \left(3 x^8+20\right)-78\right]+40\right\}+15\right)}{\phi_0^2 x^2 \left[\alpha  \left(3
   x^4-10\right) x^4+15\right]^2}\,.
\end{align}
\end{subequations}
In the case $\alpha < 9/25 $, the potential is convex everywhere, such that the second slow-roll parameter $\epsilon_2$ is always positive.  The spectral index is then red for all the modes and  therefore we can reject this case.  It is important to notice that in general,
$\epsone,\epstwo\ll1$ if $\phi\gg\Mpl$. In the situation that has originally been
considered, it is required that $\phi \ll \Mpl$ during inflation~\cite{Allahverdi:2006iq,Lyth:2006ec}. Moreover, the MSSM scenario suggests that
$\Lambda\sim\Mpl^{3/2}m_{\rm SUSY}^{5/2}$ and $\phi_0\sim\Mpl^{3/4}m_{\rm SUSY}^{1/4}$.
This implies that inflation can only efficiently take place in a fine-tuned region close to the approximate inflection point (or close to the local maximum of the potential), and that $\alpha \simeq 1$, such that the second slow-roll parameter does not take values much larger than unity during inflation.  Furthermore, during the inflationary phase, the second slow-roll parameter can only pass from negative to positive values, which is the opposite of what we look for. As a consequence, this regime is not
of interest in view of the scope of the present paper.  

Nonetheless, allowing for $\phi \gg \Mpl$ and taking $\alpha <1$ corresponds to a
potentially interesting regime in which $\epstwo$ passes from positive to negative values and then becomes again positive, while inflation is viable, since $\epsone \ll 1$.   Thus the scalar power spectrum in enhanced in the transitory phase, at the price of requiring super-Planckian field values, where the MSSM as an
effective field theory description may break down. The slow-roll parameters in this case are illustrated in FIG.~\ref{fig:pot_GMSSM}, as well as the spectral index value as a function of $\phi_k$.   
Inflation ends by violation of the slow-roll conditions when $x_{\rm end} \simeq \sqrt 2 \Mpl / \phi_0 $.   In order to obtain $\phi_k$ (the inflaton field value at horizon exit)
for each given mode, and then to determine the scalar power spectrum, we again
numerically integrate
the Klein-Gordon equation in the slow-roll approximation.   The predictions for the scalar spectral index in the 2-dimensional parameter space of the model, for the two pivot scales $\kp=0.05 \Mpc^{-1}$ and $\kd = 42 \Mpc^{-1}$, are presented in FIG.~\ref{fig:ns_GMSSM}.   One can observe that imposing the spectral index to be in the limits from the Planck experiment, 
it then follwos that the spectral index evaluated at $\kd$ is always below unity.  Thus there is no overlap between parametric regions where predictions for CMB anisotropies match
observation and regions associated with an increase of power on CMB distortion scales.  We can therefore conclude that the model does not give rise to any observable distortion of the CMB spectrum.

\subsection{Generalized Renormalizable Inflection Point Inflation (GRIPI)}

\begin{figure}
\begin{center}
\includegraphics[width=8.5cm]{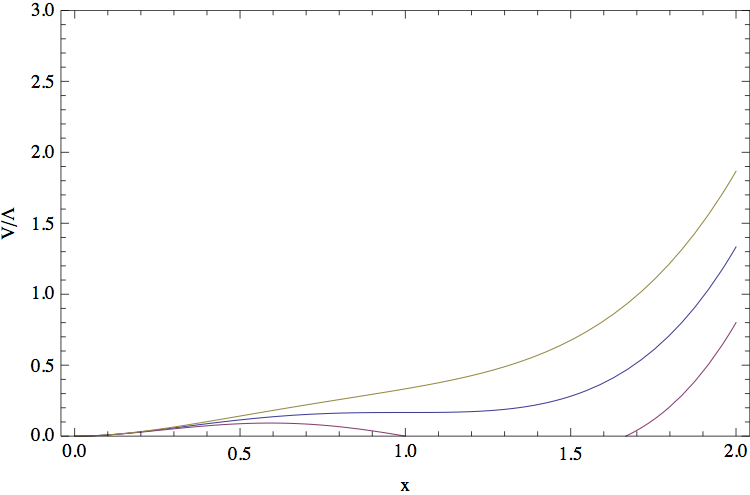}~~~~\includegraphics[width=8.5cm]{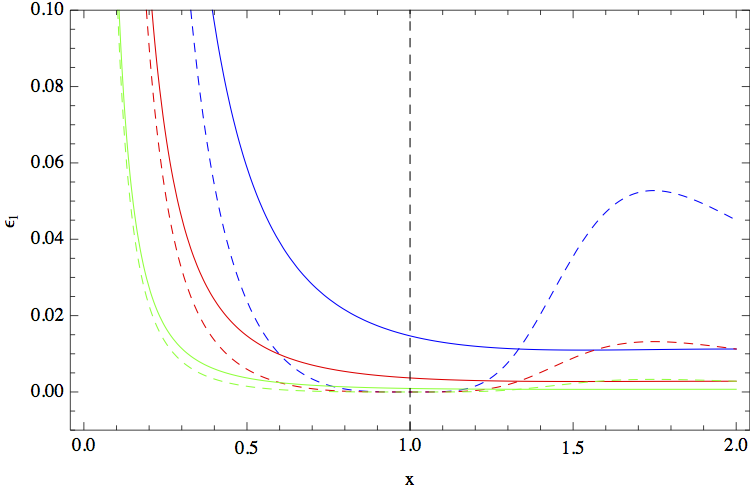}
\includegraphics[width=8.5cm]{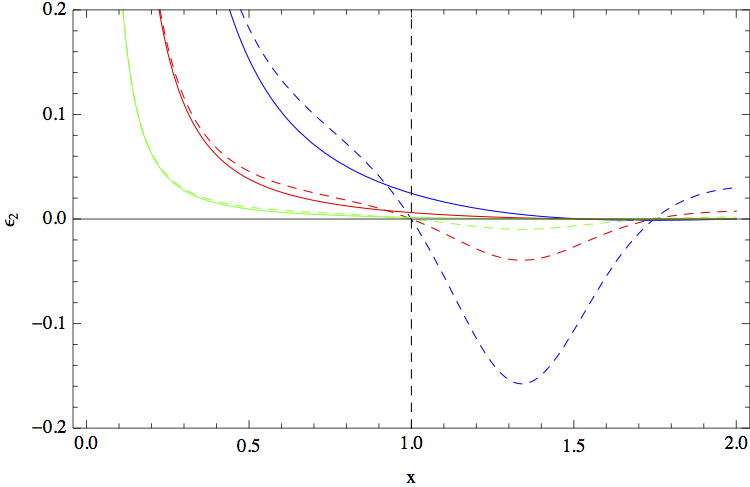}~~~~\includegraphics[width=8.5cm]{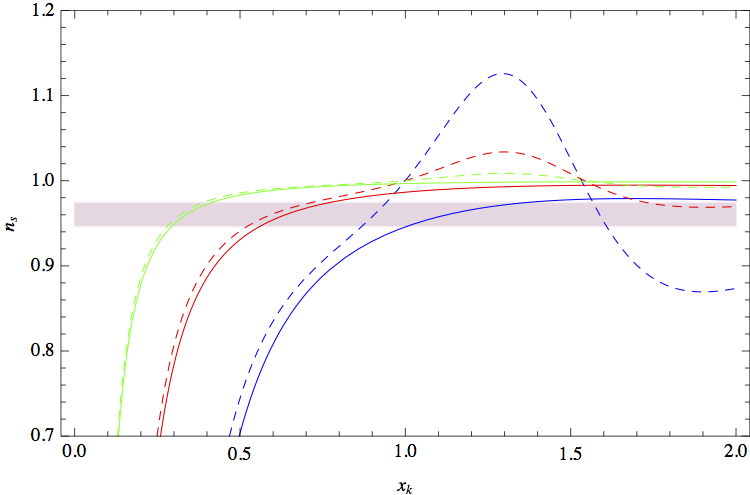}
\caption{\label{fig:pot_GIP}Field potential (top left) for the Generalized Renormalizable Inflection Point (GRIPI) model, for $\alpha = 0.8 /1 / 1.2$ (respectively in yellow, blue and red).  First and second slow-roll parameters (respectively top right and bottom left) and spectral index value  as a function of $\phi_k$ (bottom right), for the parameters $\phi_0 = 10 \Mpl$ (blue), $\phi_0 = 20 \Mpl $ (red) and $\phi_0 = 40 \Mpl$ (green), and $\alpha = 0.5 $ (solid) or $\alpha = 1$ (dashed).  As for the GMSSM model, a phase with $\ns > 1$ occurs between  two phases with $\ns < 1$ when $\alpha < 1$ and $\phi_0 \gg \Mpl $.  } 
\end{center}
\end{figure}

\begin{figure}
\begin{center}
\includegraphics[width=12.cm]{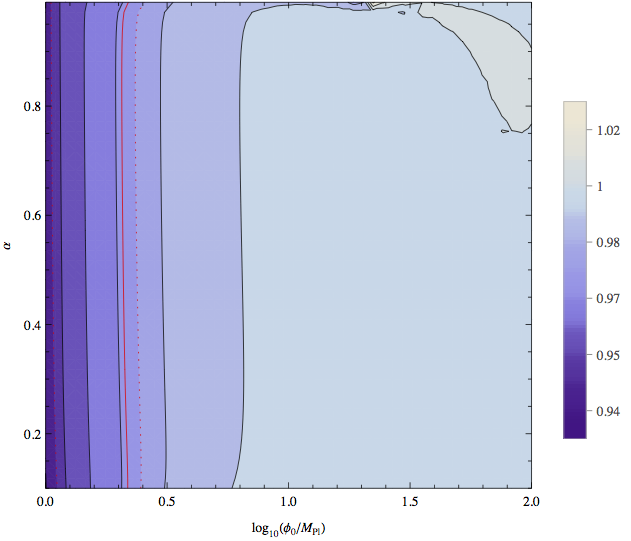}
\caption{\label{fig:ns_GIP}Spectral index for the GRIPI model evaluated for the pivot scale of CMB  distortions $k = 42 \Mpc^{-1}$ (right). To the left of the solid red contour, there is the Planck-allowed region for the spectral index evaluated
at the scale $\kp = 0.05 \ \Mpc^{-1}$.   The same contours at the scale $\kd$ (dashed red) illustrate how the spectral index changes when the scale varies.
}
\end{center}
\end{figure}

The Generalized Renormalizable Inflection Point model has features that
are very similar to the GMSSM scenario. They only differ in the
powers of the inflation field in the potential, which is now given by
\be
V(\phi) = \Lambda \left( x^ 2 - \frac 4 3 \alpha x^ 3 + \frac 1 2 \alpha x^ 4   \right)\,.
\ee
Again, we define $x \equiv \phi / \phi_0 $, and $\alpha$ is a dimensionless parameter that encodes the deviation from the Renormalizable Inflection Point (RIPI) scenario, which features an
exact inflection point where $\epsone = 0$. In the classical slow-roll approximation, this point seperates regions where $\ns > 1$ from those where $\ns < 1$ by an infinite number of e-folds,
such that the RIPI scenario is not interesting from the point of view of CMB distortions.   The theoretical motivation for the model are rather similar to the GMSSM case.
In the present case, the potential can emerge within the MSSM augmented by three additional superfields representing right handed neutrinos.  In this case the A-term is cubic, and the anticipated range for the soft SUSY-breaking mass requires $\phi_0 \sim 10^{14}$ GeV~\cite{Hotchkiss:2011am}.   The field potential is represented in FIG.~\ref{fig:pot_GIP}.  Setting $\alpha = 1$, one recovers the inflection point model.  If $\alpha > 1$, the potential develops a maximum and, as for the GMSSM model, there are three possible inflationary regimes.  If  $\alpha < 1$ the potential increases monotonously and there is no exactly flat inflection point. The slow-roll parameters read~\cite{Martin:2013tda}
\be
\epsone = \Mpl^2 \frac{72 \left[\alpha  (x-2) x+1 \right]^2}{\phi_0^2 x^2 \left[ \alpha  x (3 x-8)+6 \right]^2}\,,
\ee
\be
\epstwo = \Mpl^2 \frac{24 \left(\alpha  x \left\{ x \alpha  \left[ 3 (x-4) x+16\right] +3 x-16\right\}+6\right)}{\phi_0^2 x^2 \left[ \alpha  x (3 x-8)+6\right]^2}\,,
\ee
and together with the spectral index as function of $\phi_k$, they are presented in FIG.~\ref{fig:pot_GIP}.  For arguments identical to those in the GMSSM case, the usual regime with $\phi_0 \ll \Mpl$ cannot give rise simultaneously to a red scalar power spectrum on CMB anisotropy scales and to a blue spectrum on smaller scales.  From that perspective,
the regime of interest is where $\phi_0 \gtrsim \Mpl$ with $\alpha < 1$, but this one is
not well-motivated within the original SUSY framework, and perhaps it should
therefore be considered as a toy model.

As for the GMSSM model, the end of inflation occurs in the small field limit $\phi \ll \phi_0$ when slow-roll conditions are violated at $x_{\rm end} \simeq \sqrt 2 \Mpl / \phi_0 $.  
Then, integrating the Klein-Gordon equation for the scalar field in the slow-roll approximation, we have determined numerically the field values $\phi_k$ at which the corresponding mode $k$ exits the horizon.  We have explored the two-dimensional parameter space by calculating the spectral index predictions on the scales $\kp$ and $\kd$.  Those are represented in FIG.~\ref{fig:ns_GIP}.  Imposing the 95\% C.L. Planck constraints, we find that there is no overlap with the regions leading to $\ns >1 $ on CMB distortion scales.  
We can therefore conclude that the model does not give rise to any observable distortion of the CMB frequency spectrum.

\subsection{Running-mass Inflation (RMI)}
\label{sec:RMI}

\begin{figure}
\begin{center}
\includegraphics[width=8.5cm]{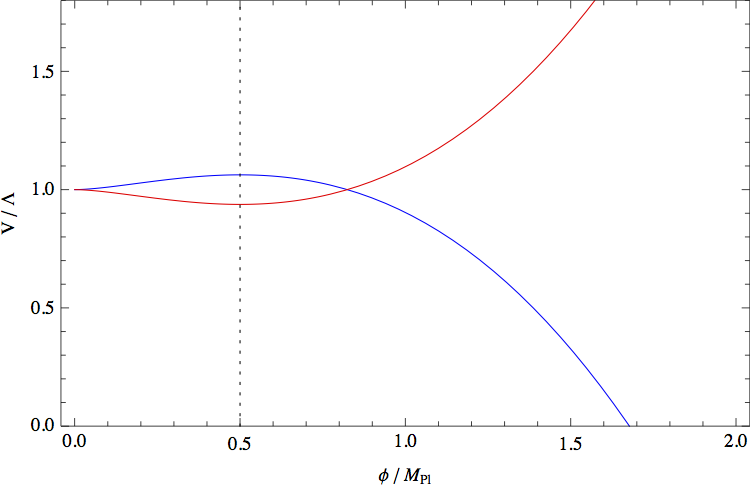}~~~~\includegraphics[width=8.5cm]{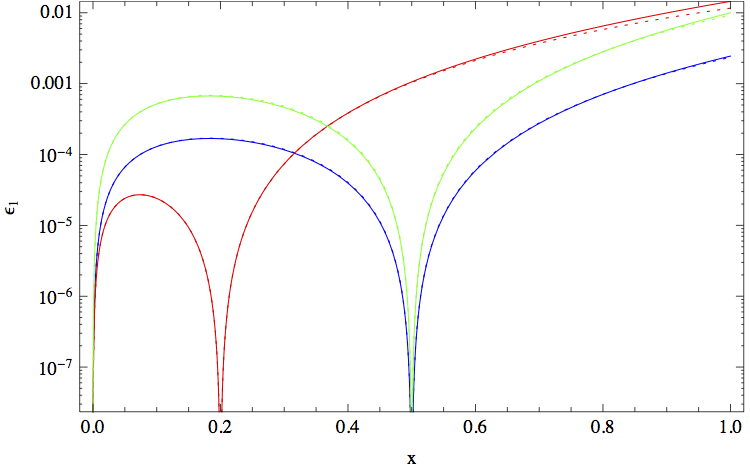}
\includegraphics[width=8.5cm]{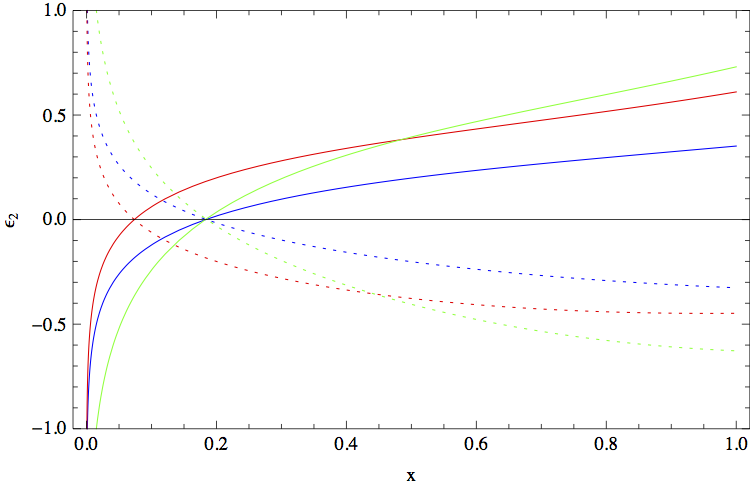}~~~~\includegraphics[width=8.5cm]{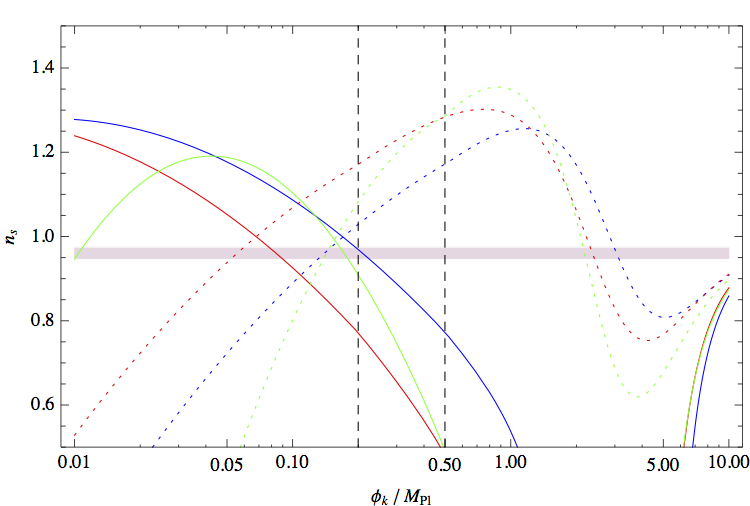}
\caption{\label{fig:pot_RMI}Field potential (top left) of the Running Mass model (RMI) for $\phi_0 = 0.5 \Mpl$,  $c = 1$ (blue) and $c = -1$ (red). First and second slow-roll parameters (respectively top right and bottom left) and spectral index as a function of $\phi_k$ (bottom right) are represented for the parameters $\phi_0 = 0.5 \Mpl$ and $c = 0.1$ (solid blue), $ \phi_0 = 0.2 \Mpl$ and $c = 0.1$ (solid red), $\phi_0 = 0.5 \Mpl $ and $c = 0.2 $ (solid green), $\phi_0 = 0.5 \Mpl$ and $c = -0.1$ (dotted blue), $ \phi_0 = 0.2 \Mpl$ and $c = -0.1$ (dotted red), $\phi_0 = 0.5 \Mpl $ and $c = -0.2 $ (dotted green).
 } 
\end{center}
\end{figure}

In this model, which assumes a supersymmectric framework, a flat direction of the potential is lifted by a soft-SUSY breaking mass term~\cite{Stewart:1996ey,Stewart:1997wg,Covi:1998jp,Covi:1998mb,Covi:2004tp}, while higher order terms are expected to be
suppressed if the inflaton field values satisfy $\phi \ll \Mpl$.   The tree level potential receives loop logarithmic corrections, which gives rise to the potential
\be
V(\phi ) = \Lambda \left[ 1 - \frac c 2 \left( - \frac 1 2 + \ln \frac{\phi}{\phi_0} \right) \frac{\phi^2}{\Mpl^2}  \right]~,
\ee
where $c$ is a dimensionless parameter that can be either positive or negative.  When assuming a
Hubble-scale soft mass $ m \simeq H$, one can estimate that $c \lesssim 0.1$.~\cite{Martin:2013tda}.  At the value  $\phi=\phi_0$, there is an extrema of the potential, a maximum if $c>0$ and a minimum in the opposite case.  The shape of the potential is shown in FIG.~\ref{fig:pot_RMI}.  Four regimes can therefore be  identified, referred to as RMI1, RMI2, RMI3 and RMI4 in Ref.~\cite{Martin:2013tda}.   RMI1 corresponds to $c >0$ and $\phi < \phi_0$ (inflation proceeds from the maximum towards small fields);  RMI2 to $c>0$ and $\phi > \phi_0$ (inflation proceeds from the maximum towards large fields);  RMI3 to $c< 0$ and $\phi < \phi_0$ (trajectories roll from small field values toward the minimum) and RMI4 corresponds to $c < 0$ and $\phi > \phi_0$ (the field rolls from large values toward the minimum of the potential).  The slow-roll parameters are given by~\cite{Martin:2013tda}
 \be
\epsone = \frac{8 c^2 x^2 \ln ^2\left(\frac{x}{\phi_0}\right)}{\left[-2 c x^2 \ln \left(\frac{x}{\phi_0}\right)+c x^2+4\right]^2}~,
  \ee
  \be
\epstwo =  \frac{8 c \left\{\ln \left(\frac{x}{\phi_0}\right) \left[2 c x^2 \ln \left(\frac{x}{\phi_0}\right)-c
   x^2+4\right]+c x^2+4\right\}}{\left[-2 c x^2  \left(\frac{x}{\phi_0}\right)+c x^2+4\right]^2}~,
  \ee
 where $x \equiv \phi / \Mpl$.  They are represented on FIG.~\ref{fig:pot_RMI} with the spectral index as a function of $\phi_k$.   
 
The RMI2 regime is not relevant in the present context, because there is no field value for which $\epstwo (\phi ) < 0$ and the spectral index is always red. In contrast, 
for RMI1, RMI3 and RMI4, it is possible to find values of $\phi_k$ that lead to a spectral index in agreement with the Planck observations and that grows larger than unity at some stage of the subsequent evolution.  The first two criteria of Sec.~\ref{sec:crit} are therefore satisfied and we have integrated numerically the slow-roll dynamics in order to determine $\phi_k$ for the two pivot scales $\kp$ and $\kd$.

 \subsubsection{RMI1:  $c > 0$, $\phi < \phi_0$}
 
In the RMI1 regime, inflation does not end due to a violation of the slow-roll conditions but when the potential develops a tachyonic waterfall instability due to the presence of some auxiliary fields\footnote{One can note that $\epsone \ll 1$ in the RMI1 regime and that the slow-roll conditions are broken when $\epstwo = 6 $.  This occurs at field values  $\phi \simeq \phi_0 \exp [- 6(1+1/2c) ]$, which is exponentially suppressed when $c\lesssim1$.  It is therefore realistic to consider that inflation ends before this point with a tachyonic instability.}.  There is therefore an additional parameter $\phic $ corresponding to the critical instability point below which we assume that inflation ends abruptly within less than one e-fold.   For practical reasons, we have define the dimensionless parameter $ c_2 \equiv \phic / \phi_0 $.   We focus on the regime $\phi_0 < \Mpl$.

\begin{figure}
\begin{center}
\includegraphics[width=12.cm]{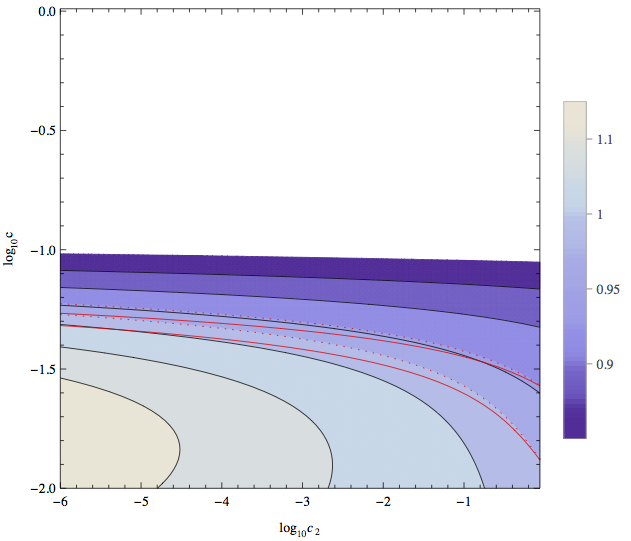}
\caption{\label{fig:ns_RMI1}Contours of the spectral index for the RMI1 regime, with $\phi_0 = 0.1 \Mpl$, evaluated for the pivot scale of CMB  distortions $k = 42 \Mpc^{-1}$ and Planck-allowed region for the spectral index evaluated at the scale $\kp = 0.05 \ \Mpc^{-1}$ (red).   The corresponding contours at the scale $\kd$ are added (dashed red) to visualize how the spectral index changes when the scale varies.  The white region correspond to values of $\phi_k$ too close of $\phi_0 $ to be numerically tractable.  In this case the spectral index is well approximated by $\ns = 1 - 2 \epsilon_k \simeq 1- 4 c $.}
\end{center}
\end{figure}

Contours of the scalar spectral index at the scale $\kd$ in the plane $(c,c_2)$  are presented in FIG.~\ref{fig:ns_RMI1} for the case $\phi_0 = 0.1 \Mpl$, as well as the Planck 95 \% C.L. limits at the pivot scale $\kp$.   Note that these contours do not change significantly when varying $\phi_0$, as long as $\phi_0 < \Mpl$.   We find no overlap between regions in agreement with the Planck observations and regions where the power spectrum is blue tilted at the scale $\kd$.   We therefore conclude that the RMI1 regime cannot lead to any detectable spectral distortion of the CMB spectrum.

 \subsubsection{RMI3:  $\beta < 0$, $\phi < \phi_0$}

In the RMI3 regime, inflation proceeds from small field values in the direction of the minimum of the potential, located at $\phi = \phi_0$.  Inflation can only stop if a waterfall instability is trigged at some point denoted as $\phi_c < \phi_0$.   For convenience, we define the additional parameter $c_2  $ here as $\phi_c = (1-c_2) \phi_0$.   

\begin{figure}
\begin{center}
\includegraphics[width=12.cm]{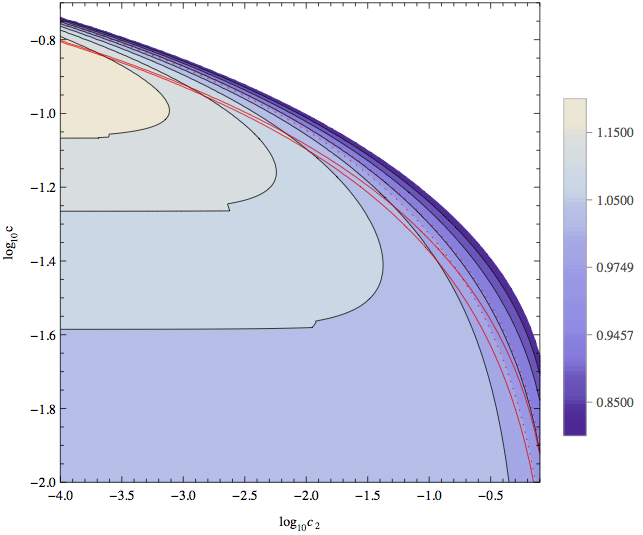}
\caption{\label{fig:ns_RMI3} Same as in FIG.~\ref{fig:ns_RMI1}, for the RMI3 regime, with $\phi_0 = 0.1 \Mpl$. }
\end{center}
\end{figure}

As for RMI1, we show in FIG.~\ref{fig:ns_RMI3} contours of the scalar spectral index at the scale $\kd$ in the plane $(c,c_2)$ for $\phi_0 = 0.1 \Mpl$, and we have superimposed the Planck 95 \% C.L. limits at the scale $\kp$.  Contrary to RMI1, we find that in a narrow region of the parameter space the Planck constraints can be satisfied together with the prediction of a blue spectrum on distortion scales.   This region corresponds to $c \sim \mathcal O(0.1) $ and $c_2 < 0.01 $, therefore requiring a certain tuning for the waterfall instability point to be located close to the minimum of the potential $\phi_0$. 
The enhancement of power is more important for smaller values of $c_2$.   The scalar power spectrum is plotted in FIG.~\ref{fig:Pzeta_RMI} for two representative parameter sets, as well as the resulting distortion spectra in FIG.~\ref{fig:distort_RMI}.  However, we find that the increase of power is not sufficient to induce an amount of spectral distortions that would be observable by PIXIE.   Note however that it could be detected by a more sensitive experiment of the PRISM-class, cf. Sec.~\ref{sec:PRISM}.  Furthermore, note that for small values of $c_2$ an increase of power on the largest CMB angular scales is also obtained.  This feature results from the abrupt variation of the spectral index in this specific region of the parameter space, which is typical of modes becoming super-Hubble during a short transition between two different regimes of the field dynamics.  Again, note that these results are independent of the value of $\phi_0$ as long as it is smaller than the Planck mass.  

 \subsubsection{RMI4:  $\beta < 0$, $\phi > \phi_0$}

\begin{figure}
\begin{center}
\includegraphics[width=8.cm]{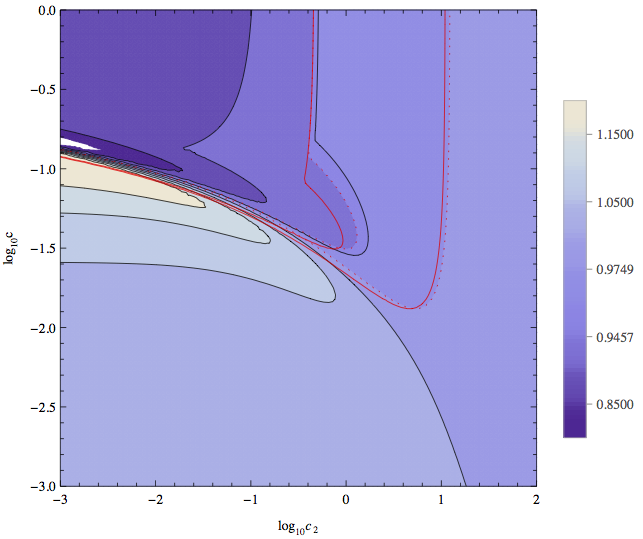} ~~~~\includegraphics[width=8.cm]{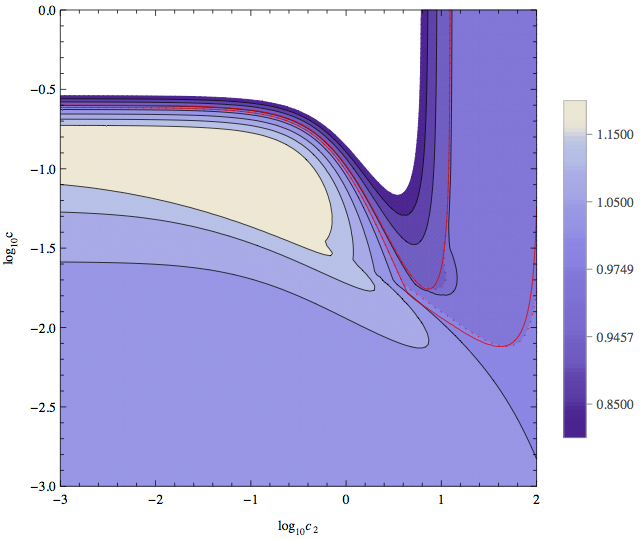}
\caption{\label{fig:ns_RMI4} Same as in FIG.~\ref{fig:ns_RMI1}, for the RMI4 regime, with $\phi_0 = \Mpl $ (left) and $\phi_0 = 0.1 \Mpl$ (right). }
\end{center}
\end{figure}

\begin{figure}
\begin{center}
\includegraphics[width=8.5cm]{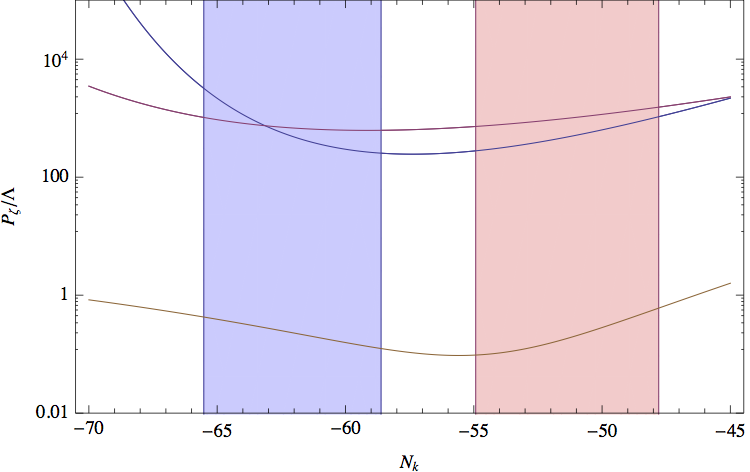}
\caption{\label{fig:Pzeta_RMI}Power spectrum of curvature perturbations for the RMI3 and RMI4 regimes with  $c_2 = 10^{-4}$, $c = - \log_{10} (-0.8)$ (RMI3, blue),  
$c_2 = \log_{10} (-2.6) $, $c = -0.1 $ (RMI3, red)   and $c_2 = 10^{-3}$, $c = - \log_{10} (-0.9)$ (RMI4, brown).  The amplitude can be normalized by a suitable choice of $\Lambda$.  }
\end{center}
\end{figure}

\begin{figure}
\begin{center}
\includegraphics[width=9.5cm]{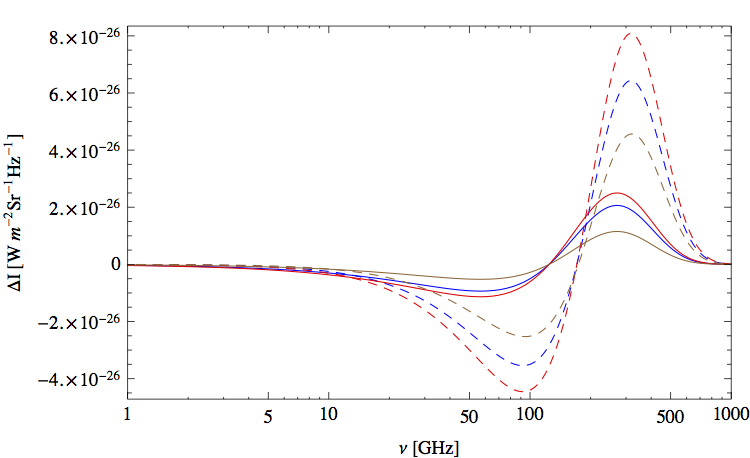}
\caption{\label{fig:distort_RMI}Spectrum of intermediate and $\mu$-distortions (respectively dashed and solid line) for the RMI model, same parameters as in FIG.~\ref{fig:Pzeta_RMI}.  They correspond to $\mu = 6.9 \times 10^{-9}$, $y = 6.4 \times 10^{-9}$ (blue), $\mu = 8.4 \times 10^{-9}$, $y = 5.2 \times 10^{-9}$ (red), $\mu = 3.9 \times 10^{-9}$, $y = 5.8 \times 10^{-9}$ (brown).    In all cases the signal is below the $2 \sigma$ sensitivity of PIXIE.  
}
\end{center}
\end{figure}

 In the RMI4 regime, inflation proceeds from large field values in the direction of the minimum of the potential located at $\phi = \phi_0$.  As for RMI3, inflation can stop only if a waterfall instability is developed before the minimum is reached.  We defined $c_2  $ such that $\phi_c = (1-c_2) \phi_0$ is the critical instability point.   In the RMI4 regime, we find that the shape of the contours of the spectral index depends on the value of $\phi_0$.  These are represented in FIG.~\ref{fig:ns_RMI4} for $\phi_0 =  \Mpl$ and $\phi_0 = 0.1 \Mpl$, and again, the Planck limits are superimposed.  When $\phi_0 \lesssim 0.1 \Mpl $ we find that regions where $\ns(\kd) >1$ do not overlap the Planck bounds.   An overlap is only possible along a very narrow band when $\phi_0 \sim \Mpl $, which therefore requires inflation to take place at field values close to the Planck mass, when supergravity corrections may destroy the flatness of the potential.   But even in that case, the increase of power (shown in FIG.~\ref{fig:Pzeta_RMI}) cannot be sufficient to 
generate observable distortions in the CMB, as illustrated in FIG.~\ref{fig:distort_RMI}.

\section{Testing multi-field inflation with CMB distortions}  \label{sec:multi_field}

In the previous Section, we have reviewed a large number of single-field models and have shown that only one of them can lead to distortions of the CMB black-body spectrum that are observable by PIXIE.  However, the WMAP experiment left room for the existence of non-gaussianity in the scalar spectrum that cannot be explained through of single-field inflation but that may arise in the context of multi-field models.  The recent Planck data has tightened the constraints on local non-gaussianities, but it still allows for $\fNL^{\rr{local}} \lesssim \mathcal O(10)$.   Multi-field inflation can arise naturally, for instance, from the string landscape.  When trajectories perform some turns in the field space, the entropic modes can mix with the adiabatic ones and act as a source for the curvature perturbations.  The field perturbations are not frozen on super-Hubble scales, contrary than in single field models.  This results not only in a potentially large level of non-gaussianity but also in a modification of the power spectrum of curvature perturbations.  In the case of an increase of power toward smaller scales, this could result in an observable amount of spectral distortions in the CMB.   In this Section, we consider three simple models of effective two-field scenarios:  i)  a softly turning trajectory,  ii)  a suddenly turning trajectory,  iii)  a waterfall trajectory.   These three scenarios have been studied in the literature~\cite{Chen:2009zp,Pi:2012gf,Gao:2013ota,Noumi:2013cfa} and analytical formulae have been derived for the curvature power spectrum.   We base our analysis on these analytical approximations and for the first time consider these models in the context of CMB distortions. 

\subsection{Softly turning trajectory}

It is possible to classify two-field trajectories in different regimes.  It is common practice to define adiabatic and isocurvature (or entropy) field components~\cite{Gordon:2000hv} by their alignment parallel or orthogonal to the field trajectory,  respectively.   Several methods can be used to derive the scalar power spectrum (as well as the corresponding  bi-spectrum and tri-spectrum), e.g. the delta N formalism~\cite{Lyth:2005fi,Sugiyama:2012tj}, the multi-field theory of linear perturbations (see Ref.~\cite{Ringeval:2007am} for a numerical implementation), the in-in formalism~\cite{Weinberg:2005vy} or effective field theory~\cite{Cheung:2007st}.  In the simplest case, when both fields are almost massless, the trajectory corresponds to one of a single field, and the curvature perturbations are directly related to the adiabatic field perturbations.   The usual single-field slow-roll formula can then be applied to the adiabatic field for calculating the scalar power spectrum, and the level of local non-gaussianity 
is small, of the order of the slow-roll parameters.   In the case where the mass of the isocurvature component is heavy ($m_{\rr{iso}} \gg H$) and the mass of the adiabatic component is sufficiently small for slow-roll inflation to take place, the trajectory is so-called \textit{effectively single field}, because isocurvature perturbations quickly decay.  In the third case, called \textit{quasi effectively single field}, the iso-curvature mass is of the order of or larger than the Hubble rate, $m_{\rr{iso}} \simeq H$ and isocurvature perturbations can have an important impact on the scalar power spectrum on a wide range of scales.

In this Section, we focus on this latter case.  We consider trajectories that initially correspond to single-field field inflation, and that at some time (referred as $t_1$) perform a turn in the field space until the time $t_2$, after which they are straight again.  During the turn, we assume for simplicity a constant angular velocity as well as a constant isocurvature mass $m_{\rr{iso}}$.  In this case, the power spectrum of curvature perturbations for modes leaving the Hubble radius during the turn is given by~\cite{Chen:2009zp,Pi:2012gf}
\be \label{eq:softturn}
\mathcal P_\zeta = \frac{H^4}{4 \pi^2 R^2 \dot \theta^2 } \left[  1 + 8 C(\mu_{\rr{iso}}) \left( \frac{\dot \theta }{H}  \right)^2  \right]\,,
\ee
where $\mu_{\rr{iso}} \equiv \sqrt{m_{\rr{iso}}^2 / H^2 - 9/4 }$, $R$ is the curvature radius of the turn in the field space, and $\theta(t)$ is the angle between the tangent vector to the trajectory and the initial direction of the trajectory.  In the regime $m_{\rr{iso}} > 3 H /2$, the function $C(\mu)$ is to a good approximation given by $C(\mu) = 1/4 \mu^2$~\cite{Pi:2012gf}.  

Strictly speaking the previous formula should be used only for constant turn trajectories and not in the case where a turn is initiated a some point within the range of observable modes and lasts for a transient period of the field evolution.  In a realistic case damped oscillations or a bump-like shape should be produced at the onset of the turn~\cite{Achucarro:2010jv,Achucarro:2012yr}, and it is not clear if such features may generate by themselves detectable distortions.  Here we assume that they can be neglected and focus on an ideal situation where the turn is sufficiently slow and lasts for a sufficient number of e-folds for Eq.~(\ref{eq:softturn}) to be a good approximation.   A full numerical calculations of spectral features during and after the turn, as well as consideration of intermediate cases interpolating from very soft to sudden turns are beyond the scope of this paper.

\begin{figure}
\begin{center}
\includegraphics[width=12.cm]{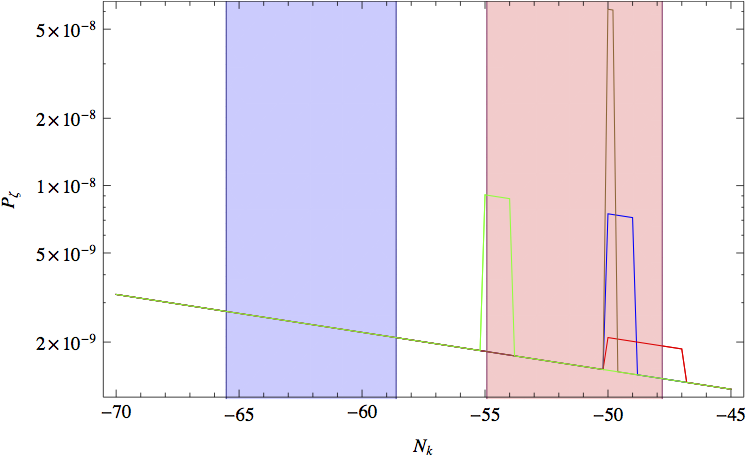}
\caption{\label{fig:soft_turn} Power spectrum of curvature perturbations for softly turning trajectories, for $\mu_{\rr{iso}} = 1$ (which gives a maximum enhancement), with $N_t = 50$ (blue, brown and red curves) and $N_t = 55$ (green), and $\Delta N_t$ corresponding to $\Delta \sigma_t = R$.  Blue and Green curves are for $\epsone / (\mu_{iso} R)^2 = 1$ and $\epsone = 0.1$, the yellow curve is for $\epsone / (\mu_{iso} R)^2 = 10$ and $\epsone = 10^{-5}$ and the red curve is for $\epsone / (\mu_{iso} R)^2 = 0.1$ and $\epsone = 10^{-3}$. }
\end{center}
\end{figure}

\begin{figure}
\begin{center}
\includegraphics[width=12.cm]{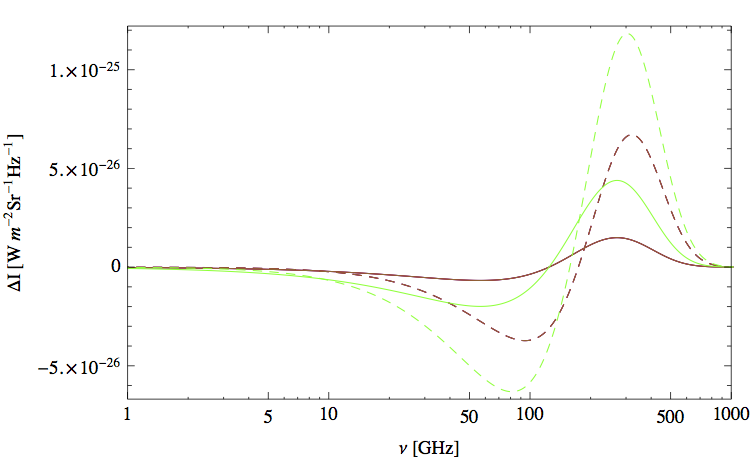}
\caption{\label{fig:soft_turn_dist} Intermediate (dashed) and $\mu$-type (solid) spectral distortion for the effective model of a softly turning trajectory and the same parameters as in FIG.~\ref{fig:soft_turn}.   The spectra for three of the parameter sets are superimposed and actually do not significantly differ from that of a single-field trajectory with constant $\ns$.  They lead to $\mu = 5.0 \times 10^{-9} $ and $y =5.4 \times 10^{-9}$.  In the case of $\mu_{\rr{iso}} = 1$ (green), which corresponds to a maximal enhancement of power, $i$-type and $\mu$-type distortions are respectively about two times or three times larger, but nevertheless should not reach the level of detectability by PIXIE ($\mu = 1.47 \times 10^{-8} $ and $y = 5.4 \times 19^{-9} $).   When the enhancement is close to maximal, the generated distortions could nevertheless be detected by a PRISM-class experiment.  }
\end{center}
\end{figure}

 At the point where the turn is initiated, $\dot \phi = R \dot \theta $, and thus one recovers the standard expression for the amplitude of the scalar spectrum.  Using the slow-roll approximation, one can therefore also relate the radius $R$ and the angular velocity $\dot \theta$ to the first slow-roll parameter $\epsone = \dot \phi^2 /(2 H^2 \Mpl^2)$, such that one obtains
 \be
 \frac{\dot \theta^2}{H^2} = \frac{\epsone}{R^2}
 \ee
Therefore, the enhancement factor goes like $ \sim \epsone \Mpl^2 / (\mu_{\rr{iso}}^2 R^2) $.   For a model with $\epsone \ll 1$ and $\mu \simeq \mathcal O(1)$, one must therefore require $R \lesssim \sqrt \epsilon \Mpl$ in order to have a significant enhancement of the power spectrum,  which is similar or smaller than the typical field variation within the time of one e-fold, $\Delta \sigma \sim \sqrt \epsilon \Mpl$.  In the same manner, if $\epsone \simeq \mathcal O(0.1) $ as for large field models, one must impose $R \lesssim \mathcal O(0.1) \Mpl $ in order to get a enhancement of power potentially leading to observable distortions in the CMB.  Therefore, in both cases, the turn occurs in typically less than one e-fold (apart in the case of spiral trajectories), which gives rise to spectral distortions not significantly different than for a trajectory with no turn.  One can therefore conclude that it will not be possible to detect the distortion signal with PIXIE in the case of softly turning trajectories.  
In order to illustrate this result, we have plotted in FIG.~\ref{fig:soft_turn} the power spectrum of curvature perturbations for several values of the parameters, of the adiabatic field variation during the turn $\Delta \sigma = R $ and of the number of e-folds $N_t$ at which the turn occurs. The resulting intermediate and $\mu$-type spectral distortions are presented in FIG.~\ref{fig:soft_turn_dist}.    It can be noticed that if the model signatures cannot be detected by PIXIE, they could nevertheless be detected by a PRISM-like experiment in the cases where the distortions are maximal.  

\subsection{Suddenly turning trajectory}

In this Section, we consider as another interesting case suddenly turning trajectories in field space.  We assume that the duration of the turn is much less than one e-fold, such that modifications of the power spectrum during the turn are non-detectable very sharp features of the spectrum.  However, as noticed in Refs~\cite{Gao:2013ota,Noumi:2013cfa}, a transversed heavy field can be excited by the sudden turn and the resulting high-frequency field oscillations subsequently affect the inflationary dynamics.  Two effects generating an imprint on the power spectrum of curvature perturbations can in general be distinguished: the modification of the Hubble parameter (called the \textit{deformation effect}) and the mixing between adiabatic and isocurvature modes (called the \textit{conversion effect}).   Interestingly, in the case of models with canonical kinetic terms, the parametric resonances induced by the two effects accidentally cancel out each other~\cite{Noumi:2013cfa}, such that the main feature generated in 
the power spectrum is a very clear peak at the turning scale.   In the following, we look for a regime in which features of the power spectrum arise on scales relevant for CMB distortions, while prior to the turn, the spectrum should remain unaffected.  For this purpose, we use the analytical approximation derived in Ref.~\cite{Noumi:2013cfa} by using the in-in formalism, 
\be \label{eq:sharpturn}
\mathcal P_\zeta (k) = \mathcal P_\zeta ^0 (k) \left[1 + \mu_{\rr{iso}} \alpha^2 \frac{(\sin x_t - x_t \cos x_t)^2}{x_t^3}  \right]~,
\ee
which is valid in the heavy mass regime $m_{\rr{iso}} \gtrsim 10 H$, and where $x_t \equiv k / k_t$.  The sudden turn is characterized by the angle $\alpha$ performed in the field space and, as before,  $\mu_{\rr{iso}} \equiv \sqrt{ m_{\rr{iso}}^2 / H^2 - 9/4  }$, as well as the scale $k_t $ that becomes super-Horizon during the turn, which occurs at the e-fold time $N_t$.   

\begin{figure}
\begin{center}
\includegraphics[width=8.cm]{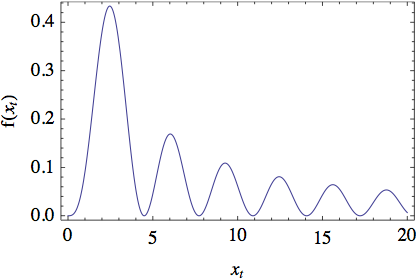}
\caption{\label{fig:fxstar} Function $f(x_t)$. }
\end{center}
\end{figure}

\begin{figure}
\begin{center}
\includegraphics[width=12.cm]{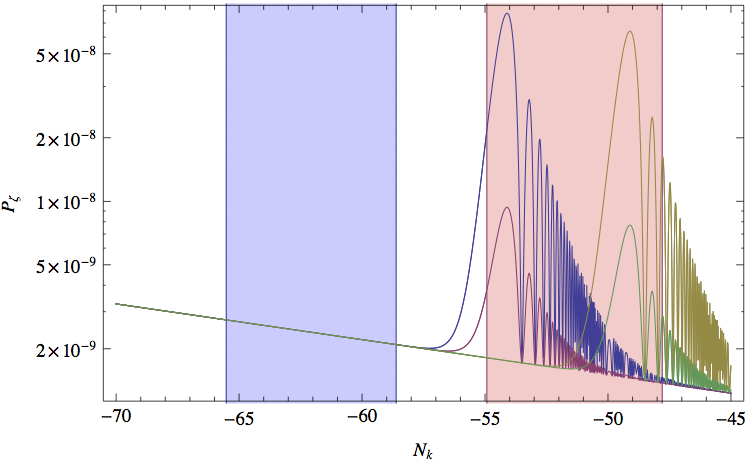}
\caption{\label{fig:sudden_turn} Power spectrum of curvature perturbations for suddenly turning trajectories, with $N_t = 55$ (blue and red curves) and $N_t = 50$ (yellow and green curves), and for $\mu_{\rr{iso}} \alpha^2 = 100 $ (blue and yellow) and $\mu_{\rr{iso}} \alpha^2 = 10 $ (red and green).  }
\end{center}
\end{figure}

Fig.~\ref{fig:fxstar} shows the function $f(x_t) \equiv (\sin x_t - x_t \cos x_t)^2 /x_t^3$, which is maximal for $x_t \simeq 2.46$ with $f(2.46) \simeq 0.43$, and then exhibits a series of damped oscillations.  The amplitude of the oscillations in the power spectrum is therefore controlled by $\mu \alpha$, whereas their frequency is a fixed prediction of the model.  Several examples have been plotted on Fig.~\ref{fig:sudden_turn}, focusing on the range a detection with CMB distortion experiments like PIXIE may be possible.   

Note that a full numerical analysis would show important deviations from Eq.~(\ref{eq:sharpturn}) after a few oscillations~\cite{Noumi:2013cfa,Gao:2012uq,Park:2012rh}, but here we argue and we have checked that the distortion signal mostly comes from the first three oscillations in the scalar power spectrum (one gets only changes at the percent level in the distortion spectrum when oscillations are cut after three or four oscillations), where our approximation only slightly overestimates the peak amplitude.  Note also that oscillations around zero compensate when integrated over wavelength modes and therefore do not lead to distortions.  

\begin{figure}
\begin{center}
\includegraphics[width=12.cm]{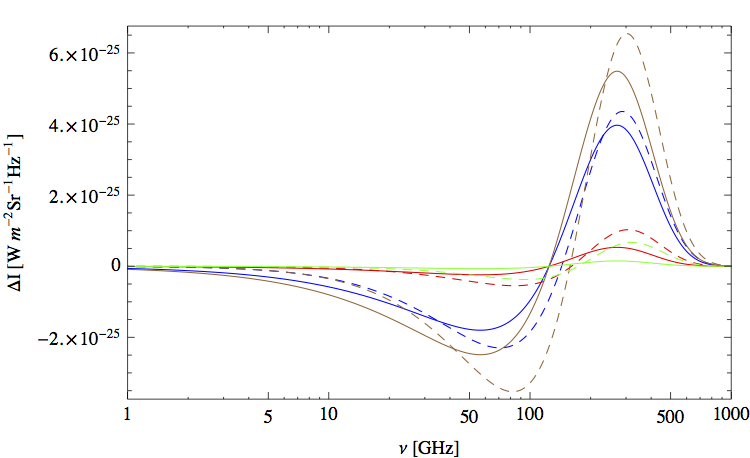}
\caption{\label{fig:sudden_turn_dist} Intermediate (dashed) and $\mu$-type (solid) spectral distortions for the effective model of a sudden-turning trajectory, and same parameter values than in FIG.~\ref{fig:sudden_turn}.   The spectra corresponding to $\mu_{\rr{iso}} \alpha^2 = 100 $ are detectable by PIXIE at more than $5 \sigma$, whereas in the two other cases they reach no more than the $2 \sigma$ level.   }
\end{center}
\end{figure}

 Because of the oscillatory shape of the spectrum, one cannot refer to Eq.~(\ref{eq:limit_PIXIE}) for bounding the scalar spectrum amplitude, since it assumes a power-law shape with spectral index close to unity.  Instead one has to refer directly to the expected sensitivity of PIXIE, given by Eq.~(\ref{PIXIE:mu:5sigma}).   The distortion spectra in Fig.~\ref{fig:sudden_turn_dist} illustrate that parameters such that $\mu_{\rr{iso}} \alpha^2 \sim 10^2 $ are detectable at $5\sigma$ confidence level, whereas $\mu_{\rr{iso}} \alpha^2 \sim10 $ does not lead to a detection at more than about $2\sigma$.  For angles $\alpha \lesssim \pi/2  $, one therefore requires 
\be \label{eq:misoconstraint}
m_{\rr{iso}} \gtrsim \mathcal O(100) H 
\ee 
for inducing an observable level of CMB distortions.  The turning scale is also required to lie within the range of observable modes. 
 Contrary to the single-field VHI model, we find that the intermediate $i$-type distortions are more important than those of the $\mu$-type, because the oscillatory features in the spectrum are damped and thus the smallest scales contribute less significantly to the distortion signal.  Note also that a variation of the turning scale changes the ratio between $i$-type and $\mu$-type distortions.  
 
We therefore conclude that sudden turning trajectories may lead to observable distortions with PIXIE if the iso-curvature mass satisfies the bound of Eq.~(\ref{eq:misoconstraint}), and if the turn takes place about 50-55 e-folds before the end of inflation.  An ultimate experiment such as PRISM would reduce the bound of Eq.~(\ref{eq:misoconstraint}) by about one order of magnitude. 

\subsection{Mild waterfall trajectory} \label{sec:hybrid2field}

Hybrid models are a particularly well-motivated class of inflation models, because they can be embedded in various high-energy frameworks. They are most commonly studied in the context of supersymmetry, where the most prominent examples are the well-known F- and D-term models~\cite{Dvali:1994ms,Binetruy:1996xj,Halyo:1996pp,Garbrecht:2006az,Clauwens:2007wc,Kallosh:2003ux}.   Moreover, they do not necessarily require super-Planckian field values (contrary to large field models), neither a fine-tuning of  initial field values~\cite{Clesse:2008pf,Clesse:2009ur,Clesse:2009zd,Easther:2013bga} (contrary to small field models), and a long phase of inflation can occur for a very low energy density.   However, for the original hybrid model~\cite{Linde:1993cn,Copeland:1994vg}, the scalar spectral index takes values larger than unity, which is now excluded by Planck, whereas for the simplest versions of F-term and D-term models, it must be larger than $\ns \gtrsim 0.98 $ which is now observationally disfavored~\footnote{The spectral index can nevertheless be lowered to 
acceptable values for the F-term model if a soft-SUSY breaking mass term is added to the potential~\cite{Pallis:2013dxa}.  This nevertheless requires some tuning of the parameters.}.   In hybrid models, inflation is usually realized along a nearly flat valley of the potential.  It ends when an auxiliary field develops a Higgs-type tachyonic instability, where a phase of tachyonic preheating is triggered~\cite{Felder:2000hj,Felder:2001kt,Copeland:2002ku}. Eventually, the field configuration reaches one of the global minima of the potential.   It is a common assumption to consider a nearly instantaneous waterfall phase (lasting less than one e-fold).   But it has been shown that inflation can continue during the waterfall for more than 60 e-folds~\cite{Clesse:2010iz,Kodama:2011vs,Clesse:2012dw,Clesse:2013jra,Abolhasani:2010kn,Mulryne:2011ni}, what consequently modifies the observable predictions of hybrid models.  It is also interesting to notice that topological defects that can be formed at the point of instability, 
and that may have dramatic consequences for cosmology are conveniently stretched outside our observable patch of the Universe.  However, for the hybrid models that have been studied so far (original and F/D-term moles), if the waterfall phase lasts for $N \gg 60$ e-folds, the scalar spectral index is given by $\ns = 1 - 4/ N_{\kp} $~\cite{Kodama:2011vs,Clesse:2012dw,Clesse:2013jra}, which is too low for being consistent with the observations made by Planck.  On the other hand, if $N \gtrsim 60$ entropic modes induce an enhancement of to the power spectrum of curvature perturbations by several orders of magnitude, well above the observed amplitude~\cite{Clesse:2013jra}.   

For these reasons, we consider in this section an intermediate case where the waterfall regime lasts for $20 \lesssim N \lesssim 60$ e-folds of inflation.   The model is based 
on the two-field potential
\be
V(\phi, \psi) = \Lambda \left[  \left( 1 - \frac{\psi^2}{M^2} \right)^2 +  \frac{\phi}{\mu}  + \frac{2 \phi^2 \psi^2}{\phic^2 M^2} \right] ~,
\ee
where $\psi = \pm M $ is the position of the global minima of the potential at $\phi = 0$, and where the parameter $\mu$ controls the slope of the potential close to the critical point of instability  $\phic$.   The observable modes with CMB anisotropies leave the Hubble horizon prior to the inflaton reaching the instability point, and thus the scalar power spectrum on those scales can be calculated using the standard single-field slow-roll formalism.    If there is no additional term to the potential, one would get $\epstwo =0$, and because a mild waterfall phase requires $\epsilon_1 \ll 1$ at the point of instability, the scalar spectral index would be very close to unity.   This problem can be avoided if the inflaton $\phi$ gets a mass term driving $\epstwo$ to acceptable values.   Such an additional mass term could arise for instance due to logarithmic loop corrections to the potential \footnote{Notice however that for F-term and D-term, the mass of $\phi$ at the critical point in the mild waterfall regime cannot be reconciled with the 
spectral index.  Other models should therefore be envisaged}.    

According to Refs.~\cite{Kodama:2011vs,Clesse:2012dw,Clesse:2013jra}, two phases can be distinguished during the waterfall.  During the \textit{phase-1}, the transverse field contribution to the slope of the potential in the inflaton direction is subdominant compared to the $\sim \mu^{-1}$ term.  It is dominant during the \textit{phase-2}, where the dynamics is effectively of the single-field type.   During \textit{phase-1},  the power spectrum is enhanced due to the important contribution from entropic modes.   In the following, we search for a region of the parameter space leading to an increase of power by more than a factor ten (which corresponds approximatively to the level of detectability of the resulting CMB distortions) on CMB distortions scales, compared to the amplitude on CMB anisotropy scales.  

An analytical approximation of the power spectrum of curvature perturbations for modes exiting the Hubble radius in the \textit{phase-1} has been derived in Ref.~\cite{Clesse:2013jra} by using the $\delta N$ formalism, 
\be \label{eq:Pzeta_exit_in_phase1}
{\cal P}_\zeta (k) \simeq \frac{\Lambda M^2 \mu \phic}{192 \pi^2  \Mpl^6 \chi_2 \psi_k^2}  
\,,
\ee
where $\xi $ and $\chi$ are defined as $\phi = \phic \exp (\xi) \simeq \phic (1+\xi) $ and $\psi = \psi_0 \exp (\chi)$.   The validity of this approximation has been checked numerically by using the $\delta N $ formalism and by integrating the exact multi-field background plus linear perturbation dynamics.   In different patches of the Universe a non vanishing value $\psi_0$ of the auxiliary field at the critical instability point is expected due to quantum stochastic fluctuations.  Solving the corresponding Langevin equation for the auxiliary field gives 
\be \label{eq:psi0}
 \sqrt{\langle \psi_0^2 \rangle} = \left( \frac{\Lambda  \sqrt{2 \phic \mu_1} M}{96 \pi^{3/2} }   \right)^{1/2}~.
\ee  
One must therefore interpret the classical waterfall regime as quickly emerging from a stochastic patch where the quantum fluctuations of $\psi $ are important.  The subscript $_2$ in Eq.~\ref{eq:Pzeta_exit_in_phase1} denotes an evaluation at the time of transition between \textit{phase-1} and \textit{phase-2}.  Solving the slow-roll dynamics in these phases, one gets 
\be
\chi_2 \equiv  \ln  \left( \frac{\phic^{1/2} M}{ 2  \mu_1^{1/2} \psi_{0}}  \right)~.
\ee
as well as the relation  $\chi_k = 4 \phic \mu_1 \xi_k^2 / M^2$ for modes exiting the Hubble horizon during \textit{phase-1}, with
$\xi_k = - \Mpl^2 (N_k + N_2 ) / (\mu_1 \phic )$.   The number of e-folds $N_2$ realized in \textit{phase-2} is well approximated by
\be
N_2 = - \frac{M^2}{8 \Mpl^2 \sqrt{C}}  \left[  \rr{arctan} \left( \frac{\xi_{\rr{end}}}{\sqrt C}  \right) - \rr{arctan} \left( \frac{\xi_{2}}{\sqrt C}  \right)   \right]~,
\ee
with $C \equiv -xi_2 ^2 + \psi_0^2 \exp [2 \chi_2]/(2 \phic^2) $ and $\xi_{\rr{end}} = - M^2 / (8 \Mpl^2) $.

\begin{figure}
\begin{center}
\includegraphics[width=12.cm]{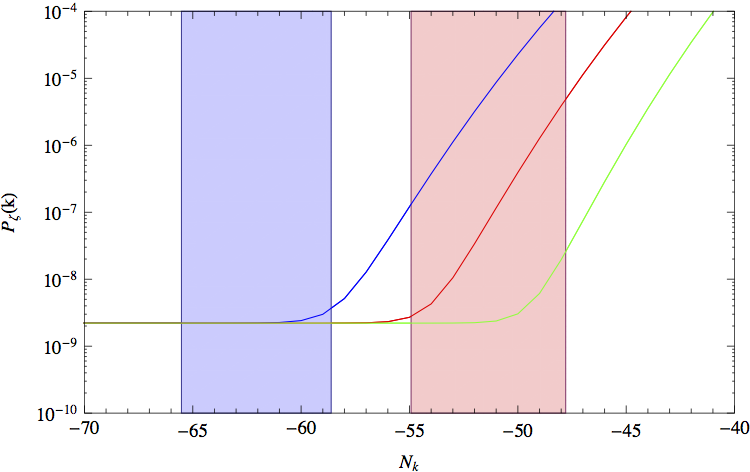}
\caption{\label{fig:spectrum waterfall} Power spectrum of curvature perturbations for the waterfall inflation model, with $\Pi = 250 / 300 /350$ (respectively green, red and blue lines). }
\end{center}
\end{figure}

\begin{figure}
\begin{center}
\includegraphics[width=8.cm]{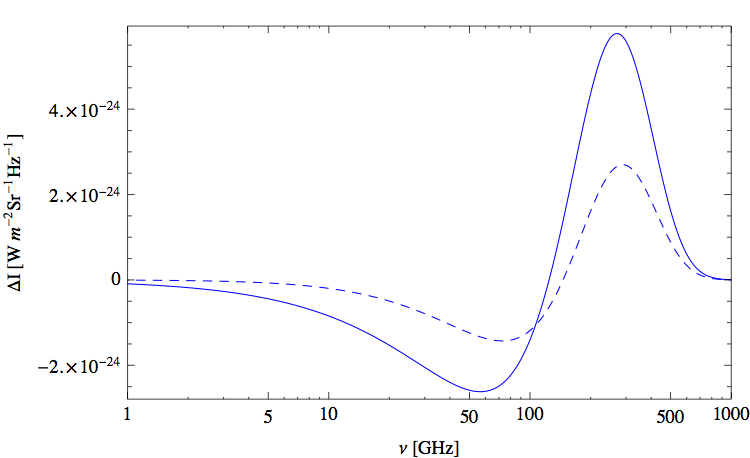}  ~~~~ \includegraphics[width=8.cm]{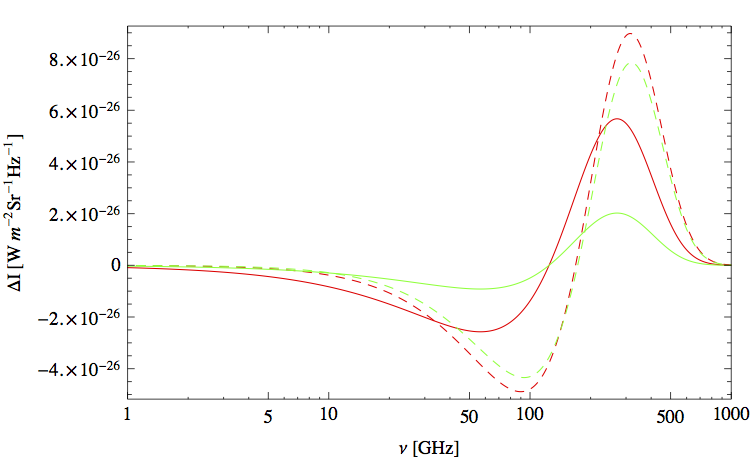}
\caption{\label{fig:waterfall_dist} Intermediate (dashed) and $\mu$-type (solid) spectral distortion for the mild waterfall model, with $\Pi = 350$ (left panel), $\Pi = 300$ (right panel, red line) and $\Pi = 250 $ (right panel, green line).   In the first case the signal cannot be detected, whereas the latter two cases are detectable by PIXIE at about $2\sigma$ level. These parameter sets induce respectively $\mu$-type and $y$-type distortions at the level $\mu  = 1.93 \times 10^{-6}, 1.90 \times 10^{-8}, 6.79 \times 10^{-9} $ and $y =( 6.64 , 5.32 , 5.32) \times 10^{-9} $. }
\end{center}
\end{figure}

All the ingredients are now given to calculate the power spectrum of curvature perturbations.  Notice that Eq.~(\ref{eq:Pzeta_exit_in_phase1}) can be used also for the modes which exit the horizon a few e-folds before the instability point is reached and which are affected on super-horizon scales by entropy perturbations generated during the first phase of the waterfall.   Another important remark is that the power spectrum is fixed by the combination of the potential parameters $\Pi \equiv M (\phic \mu_1)^{1/2} /\Mpl^2$.  The power spectrum of curvature perturbation has been plotted in Fig.~\ref{fig:spectrum waterfall} for different values of $\Pi$.  We find that within the range
\be
250 \lesssim \Pi^2 \lesssim 350
\ee
one obtains a significant increase of power on CMB distortion scales whereas the spectrum amplitude remains nearly scale invariant on CMB anisotropy scales.   The resulting spectral distortions are shown in FIG.~\ref{fig:waterfall_dist}.    Within this range, given the bound we have determined in Eq.~(\ref{eq:limit_PIXIE}), as well as the expected sensitivity for PIXIE given by Eq.~(\ref{PIXIE:mu:5sigma}), this enhancement corresponds to a detectable signal at about $2\sigma $ of confidence level.  
 
\section{Testing a curvaton model with CMB distortions}  \label{sec:curvaton}

Another rather apparent way to generate a scalar power spectrum, that is red on
the largest observable angular scales, and nonetheless enhanced on the smaller
scales that are relevant for spectral distortions, is to superimpose a red with
a blue spectrum. Blue spectra are a generic prediction for simple models with a
massive curvaton, where the small scales are enhanced because these
suffer less damping after horizon exit than the larger scales. In order to achieve
a distortion signal, that complies with the observational bounds from the
Planck and WMAP experiments, we must require that the curvaton mass is of
order (but somewhat below) the Hubble rate and that the amplitude of the
curvaton spectrum is subdominant compared to the red component of
the power spectrum at the scale
$\kp=0.05 {\rm Mpc}^{-1}$, but it should be dominant for $\kd=42 {\rm Mpc}^{-1}$.
While this choice of the relative amplitudes entails a coincidence
that may be hard to motivate, the coincidence of the curvaton mass with
the Hubble rate may be explained by a mechanism that generates a
dynamic, Hubble-induced mass.

As a concrete realisation of the scenario outlined above,
we consider a massive curvaton $\sigma$. The normalised
solutions for the momentum modes are given by
\begin{align}
\sigma(\mathbf k_{\rm c},\eta)=-H_*\eta\sqrt{\frac{\pi \eta}{4}}
H_{\nu}^{(1)}\left(-|\mathbf k_{\rm c}|\eta\right)\,,
\end{align}
where
\begin{align}
\nu=\sqrt{\frac94-\frac{m^2_\sigma}{H_*^2}}\,.
\end{align}
Here, $\mathbf k_{\rm c}$ is a comoving momentum, that is
related to the physical momentum as $\mathbf k=\mathbf k_{\rm c}/a$, and
during inflation, the scale factor is $a=-1/(H\eta)$, where
$\eta$ is the conformal time. Moreover, $H_*$ is the Hubble rate
during inflation, which we take here to be constant for simplicity, and as it
is a good approximation for small-field models.
The power spectrum at the end of inflation then is
\begin{align}
{\cal P}_{\delta\sigma}(\mathbf k,t_{\rm end})=\frac{|\mathbf k_{\rm c}|^3}{2\pi^2}\sigma^2(\mathbf k_{\rm c},\eta_{\rm end})
=\frac{|\mathbf k|^3}{8\pi H} \left|H_\nu^{(1)}(|\mathbf k|/H)\right|^2
\approx \frac{H^2}{4\pi^2}\left(|\mathbf k|/H\right)^{3-2\nu}\,,
\end{align}
where the variable $t$ with or without subscript refers to comoving time.
The time at which inflation ends is denoted by $t_{\rm end}$.
Before the oscillations of the curvaton field start, the power spectrum
suffers an additional damping, such that
\begin{align}
\label{damp:Psigma}
{\cal P}_{\delta\sigma}(\mathbf k,t)
={\cal P}_{\delta\sigma}(\mathbf k,t_{\rm end})
{\rm e}^{-\int\limits_{t_{\rm end}}^{t} dt\frac{2 m^2}{3H}}
\,,\quad\textnormal{for}\quad t_{\rm end}<t<t_{\rm osc}\,.
\end{align}
We denote here by
by $t_{\rm osc}$ the time where $H\approx m_\sigma$
and curvaton oscillations start.
Similarly, for the background curvaton,
\begin{align}
\label{damp:sigma}
\sigma(t)=\sigma(t_{\rm end}){\rm e}^{-\int\limits_{t_{\rm end}}^{t} dt\frac{m^2}{3H}}
\,,\quad\textnormal{for}\quad t_{\rm end}<t<t_{\rm osc}
\,.
\end{align}
In order to obtain the curvature perturbation, we
calculate the evolution of the
cosmic fluid components from a time $t_{\rm osc}$, when the curvaton
energy density is negligible compared to the critical density, up
to a constant energy hypersurface around the
time $t_{\rm dec}$, where the curvaton decays (in sudden decay
approximation) and it makes up a fraction $\Omega_{\sigma{\rm dec}}$
of the critical density. The number of ${\rm e}$-folds depends on the
intial value of the curvaton field at $t_{\rm osc}$. One can therefore
use the $\delta N$ formalism in order to
derive the power spectrum of curvature
perturbations~\cite{Lyth:2005fi,Lyth:2006gd}
\begin{align}
{\cal P}_{\zeta_\sigma}(\mathbf k)=&{\cal P}_{\delta\sigma}(\mathbf k,t_{\rm osc})\frac49\Omega_{\sigma{\rm dec}}^2\left(\frac{1}{\sigma(t_{\rm osc})}\right)^2
={\cal P}_{\delta\sigma}(\mathbf k,t_{\rm end})\frac49\Omega_{\sigma{\rm dec}}^2\left(\frac{1}{\sigma(t_{\rm end})}\right)^2
\\\notag
\approx& a_\sigma \left(\frac{|\mathbf k|}{k_{\rm pivot}}\right)^{n_{\rm s}^{(\sigma)}-1}\,,
\end{align}
where
\begin{align}
\label{a:sigma}
a_\sigma=\frac{H_*^2}{9\pi^2\sigma^2(t_{\rm end})}\Omega^2_{\sigma{\rm dec}}\left(\frac{k_{\rm pivot}}{H_*}\right)^{3-2\nu}\,.
\end{align}
We observe that within the final expression for the curvaton-induced curvature power spectrum, the damping of the curvaton field after inflation
[Eqs.~(\ref{damp:Psigma},\ref{damp:sigma})] cancels.
Moreover, in order to obtain a strong blue tilt, which is our present phenomenological
interest, we require that $H_*\gtrsim m_\sigma$, such that
the oscillations begin very soon after the end of inflation and that this
epoch of damping plays no quantitatively
important role.

This perturbation from the curvaton
adds linearly to an extra contribution, that may be
generated by the inflaton and which we parametrise by the standard form
\begin{align}
{\cal P}_{\zeta{\rm inf}}(\mathbf k)=a_{\rm inf}\left(\frac{|\mathbf k|}{k_{\rm pivot}}\right)^{n_{\rm s}^{({\rm inf})}-1}\,,
\end{align}
such that
\begin{align}
{\cal P}_{\zeta}(\mathbf k)={\cal P}_{\zeta{\rm inf}}(\mathbf k)+{\cal P}_{\zeta_\sigma}(\mathbf k)\,.
\end{align}

We may perform phenomenological studies of these spectra
that are parametrised by the amplitudes and their tilts. In order to
interpret the parameter $a_\sigma$, we should note that we cannot
too deliberately adjust it by choosing a small $\Omega_{\sigma{\rm dec}}$,
as this is tightly bound by observational limits on non-Gaussianities
through the relation~\cite{Lyth:2006gd}
$\frac35 f_{\rm NL}=\frac34({\cal P}_{\zeta\sigma}/{\cal P}_\zeta)^2/\Omega_{\sigma{\rm dec}}$\,. Rather, one can tune $a_\sigma$ by adjusting
$\sigma(t_{\rm end})$ through initial conditions, or it may
be fixed along an approximately flat direction that is lifted due
to Hubble-induced mass terms during
inflation~\cite{Dine:1995kz,Garbrecht:2006aw}.
When neglecting the spectral tilt in Eq.~(\ref{a:sigma}), one may
conclude that $\sigma(t_{\rm end})\sim 10^4 H_*$ in order to
obtain a substantial contribution of the curvaton to the observed
anisotropies. However, the
tilt can easily be strong enough such that smaller values of
$\sigma(t_{\rm end})$ can be used in order to have
an observationally relevant
amplitude at the pivot scale. One should check in such
a situation, that the spectrum, which is then strongly enhanced on small 
scales, complies with the bounds from ultracompact minihaloes (UCMHs)~\cite{Bringmann:2011ut} and
primordial black holes (PBHs)~\cite{Josan:2009qn}.

The relevant constraint from UCMHs is~\cite{Bringmann:2011ut}
\begin{align}
\label{constraint:UCMH}
{\cal P}_\zeta(|\mathbf k_{\rm UCMH}|)<{\cal P}_{\rm UCMH}\,,
\end{align}
where
\begin{align}
{\cal P}_{\rm UCMH}\approx 5\times 10^{-8}\,\quad\textnormal{and}\quad
|\mathbf k|_{\rm UCMH}\approx 5\times 10^7 {\rm Mpc}^{-1}\,.
\end{align}
In terms of e-folds after exit of the pivot scale, $|\mathbf k|_{\rm UCMH}$
corresponds to $N_{\rm e}^{\rm UCMH}\approx21$.
Constraints from UCMHs are also present on larger scales, but as we assume that
around $|\mathbf k|\approx|\mathbf k_{\rm UCMH}|$,
${\cal P_\zeta}$ is dominated by the blue-tilted curvaton contributions,
these are fulfilled provided the relation~(\ref{constraint:UCMH})
is satisfied.

Similarly, PBHs require that~\cite{Josan:2009qn}
\begin{align}
\label{constraint:PBH}
{\cal P}_\zeta(|\mathbf k|)<{\cal P}_{\rm PBH}\,,\quad\textnormal{where}\quad
{\cal P}_{\rm PBH}\approx5\times 10^{-2}\,.
\end{align}
This constraint should hold down to the smallest scales that exit
the horizon during inflation. Again, due to the blue-tilted
component that dominates the spectrum on small scales, the PBH
constraints are most severe on the smallest scales. For definiteness,
we assume that these exit the horizon $N_{\rm e}^{\rm PBH}\approx50$
${\rm e}$-folds after the pivot scale.

\begin{figure}[t!]
\begin{center}
\epsfig{file=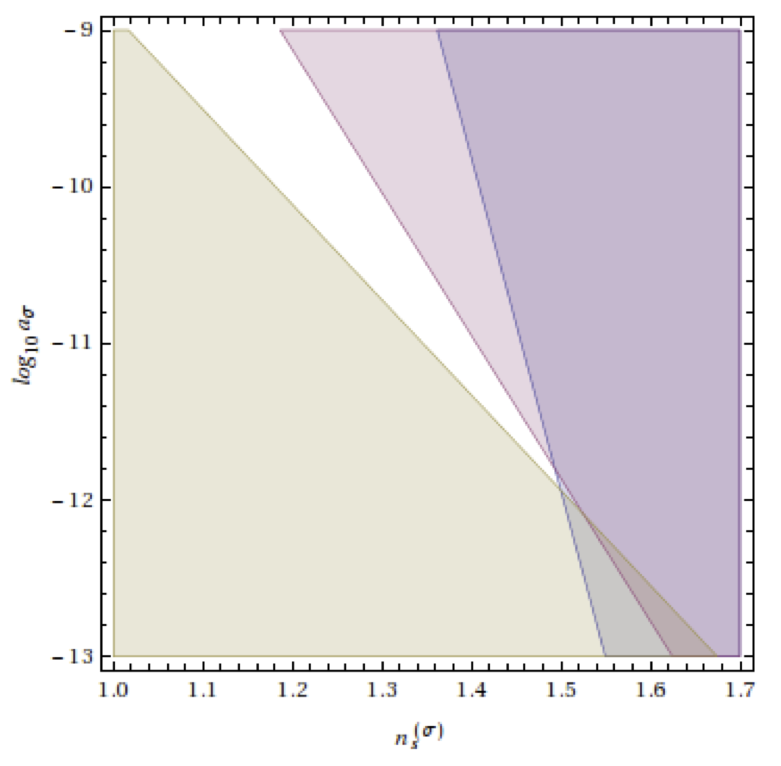,scale=0.6}
\end{center}
\caption{
\label{fig:PBHUCMH}
Constraints on $a_\sigma$ and $n^{(\sigma)}_{\rm s}$. The blue region
is excluded due to the non-observation of PBHs,
relation~(\ref{constraint:PBH}), the red region is excluded due to
constraints from UCMHs, relation~(\ref{constraint:UCMH}).
Above the brown region, the power spectrum from the curvaton
is comparable or larger than the red-tilted component of the
power spectrum at the smallest scales relevant for CMB distortions,
which is more precisely stated by relation~(\ref{P:comparable}).
The white region thus indicates the parameter space that is of
interest in view of future observations of CMB distortions.
}
\end{figure}

\begin{figure}
\begin{center}
\includegraphics[width=12.cm]{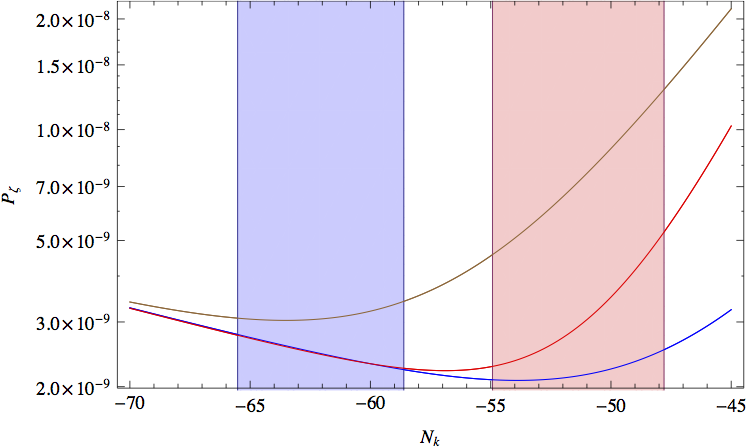}
\caption{\label{fig:curvaton} Power spectrum of curvature perturbations for the curvaton model and three sets of parameters in the white region of FIG.~\ref{fig:PBHUCMH}:  $\log_{10} a_\sigma = -10$, $\ns ^\sigma = 1.2$ (blue), $\log_{10} a_\sigma = -10$, $\ns ^\sigma = 1.3$ (red) and $ \log_{10} a_\sigma = -9$, $\ns ^\sigma = 1.2$ (brown).   }
\end{center}
\end{figure}

\begin{figure}
\begin{center}
\includegraphics[width=12.cm]{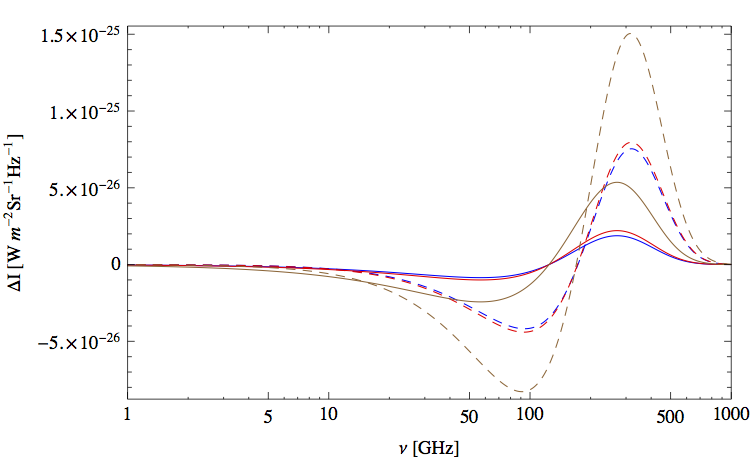}
\caption{\label{fig:curvaton_dist} Intermediate (dashed) and $\mu$-type (solid) spectral distortion for the curvaton model and same parameter values as in FIG.~\ref{fig:curvaton}.  }
\end{center}
\end{figure}

In Figure~\ref{fig:PBHUCMH}, we summarise the constraints~(\ref{constraint:UCMH},\ref{constraint:PBH}) . We also indicate the region, for which
\begin{align}
\label{P:comparable}
{\cal P}_{\zeta_\sigma}={\cal P}_{\zeta{\rm inf}}
\end{align}
(assuming
$n_{\rm s}^{({\rm inf})}=0.96$)
at $N_{\rm e}\approx14$
${\rm e}$-folds after the exit of the pivot scale, which is the smallest 
scale relevant for CMB distortions.   In FIG.~\ref{fig:curvaton} we show the total power spectrum for three parameter sets in the region allowed by UCMH and the non observation of PBHs.  The resulting spectral distortions are plotted in FIG.~\ref{fig:curvaton_dist}.  While being close, these do not reach the level of detectability, apart for parameter sets where there is a significant increase of power already on CMB angular scales, which is excluded by Planck.  One can therefore conclude that the model should not lead to detectable distortions by PIXIE when the constraints from UCMHs are satisfied.     Nevertheless, as also found for a few inflation models, a PRISM-class experiment with about ten times better sensitivity can in principle establish new constraints on the curvaton model.

\section{Forecasts for a PRISM-class probe}
\label{sec:PRISM}

Within the previous Sections, in view of the proposed sensitivity of PIXIE, we have identified
models that lead to a strongly increased power of scalar perturbations on distortion scales,
a requirement that is more precisely formulated within the criteria of Sec.~\ref{sec:crit} and
that gives rise to a rather restricted list of models. In contrast,
experiments with an improved sensitivity
should be able to probe a much wider set of inflationary models with an anomalous behaviour
of the power spectrum on smaller scales, by which we mean a deviation from spectra that
can be well descibed by a small running $|n_{\rm run}|\lsim 10^{-3}$ of the scalar spectral
$n_s$, where $n_{\rm run}$ is defined as
\begin{align}
n_{\rm run}=\frac{d \log n_s}{d \log k}\,.
\end{align}
For most viable single-field slow-roll models,
$n_{\rm run}$
is predicted to be of second order in the slow-roll parameters, i.e. $\sim-1/N_{k_{\rm p}}^2$, cf. Ref.~\cite{Martin:2013tda}.

\begin{figure}
\begin{center}
\includegraphics[width=10cm]{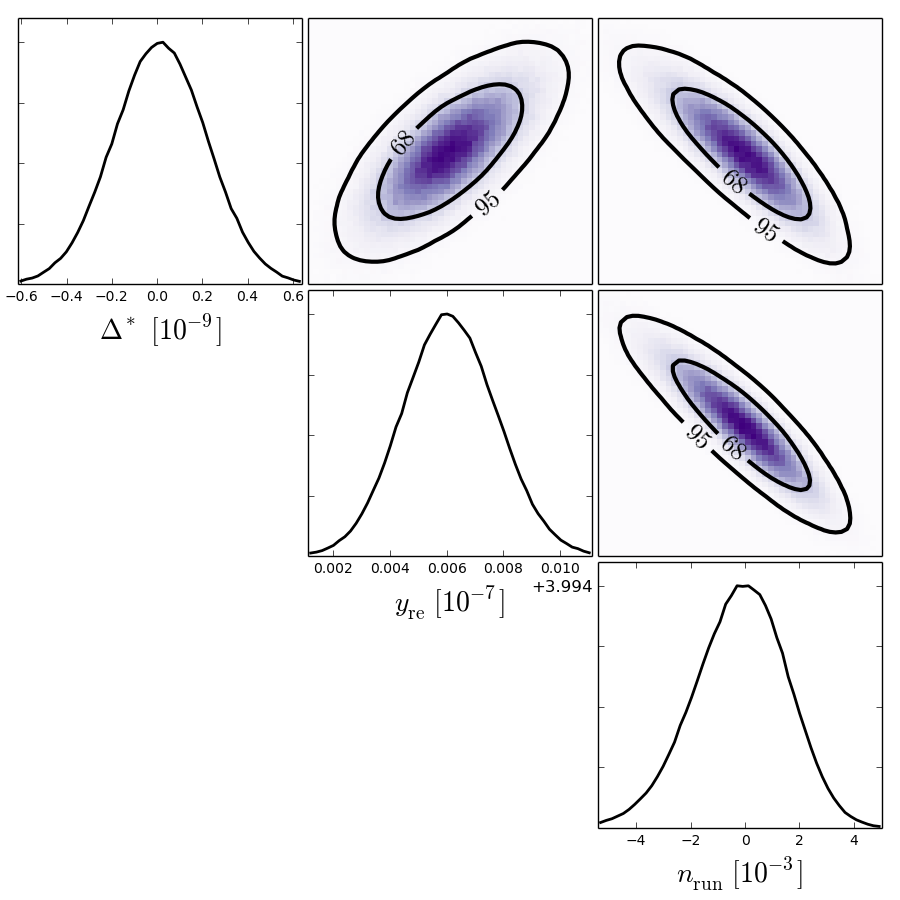}
\end{center}
\caption{\label{FIG:MCMC:diss:runns}
Projected exclusion bounds on $n_{\rm run}$ for a PRISM-class experiment with 
a sensitivity ten
times better than in Eq.~(\ref{sensitivity:PIXIE}).}
\end{figure}

For scalar power spectra that can be parametrised
in terms of $n_{\rm run}$, we present
in FIG.~\ref{FIG:MCMC:diss:runns}
a projection for the observation of an anomalously large value of $|n_{\rm run}|$.
For the fiducial model that is given by the Planck central values~(\ref{values:fiducial}),
$n_{\rm run}=0$, $y_{\rm re}=4\times 10^{-7}$ and $\Delta^*=0$,
we may expect that a PRISM-class experiment, with a sensitivity
improved by a factor of ten compared to PIXIE [cf.
Eq.~(\ref{sensitivity:PIXIE})], can exclude
$|n_{\rm run}|\gsim 4\times 10^{-3}$ at 95\% c.l. We have used the
package \Greens~\cite{Chluba:2013vsa} to obtain these results.
Similar conclusions have been drawn
in Ref.~\cite{Chluba:2013pya}.

The projections for the measurements of $n_{\rm run}$ can directly be applied to the RMI
model in the approximation where the contribution from the $\epsilon_{1k}$ term to
${\cal P}_\zeta(k)$ is neglected, which applies when during inflation, the
field $\phi$ takes sub-Planckian expectation values only. In that case, it is useful
to introduce the parameter $s=c\log(\phi_0/\phi_*)|_{k=k_{\rm p}}$, following
Ref.~\cite{Covi:2004tp}.
Its relation to the parametrisations in Sec.~\ref{sec:RMI} is given by
\begin{subequations}
\begin{align}
s=&\mp c{\rm e}^{-c N_{k_{\rm kp}}}\log c_2\,,\quad \textnormal{for RMI1 and RMI2}\,,\\
s=&- c{\rm e}^{-c N_{k_{\rm kp}}}\log (1-c_2)\,,\quad \textnormal{for RMI3 and RMI4}\,.
\end{align}
\end{subequations}
(Note that for RMI1,2,3, $c_2>0$, while for RMI4, $c_2<0$.) Using the simple relations
$n_s=-2c+2s$ and $n_{\rm run}=2sc$, the bounds from FIG.~\ref{FIG:MCMC:diss:runns} can
therefore be related to the model parameters of the RMI scenarios, see e.g.
Ref.~\cite{Chluba:2012we} for earlier projections not based on MCMC analysis.

\begin{figure}
\begin{center}
\includegraphics[width=6cm]{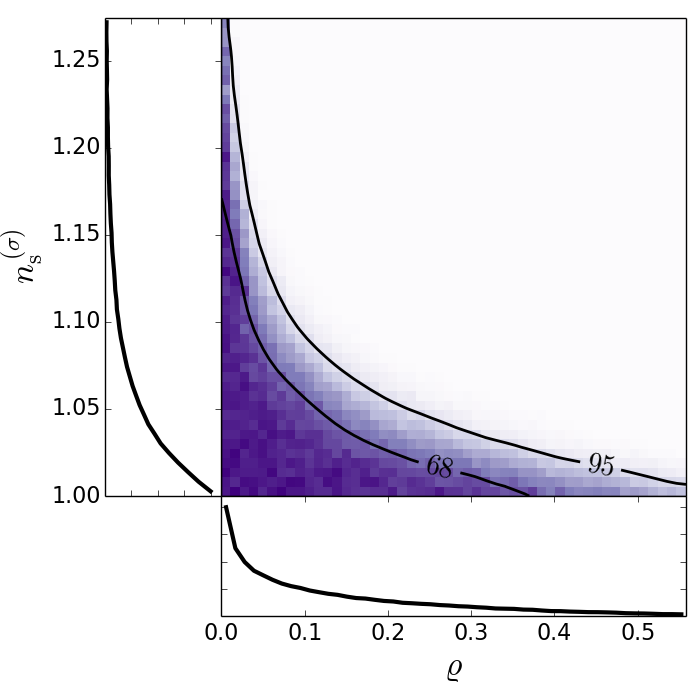}
\end{center}
\caption{\label{FIG:MCMC:curvaton}
Projected exclusion bounds on the curvaton scenario
from a PRISM-class experiment.}
\end{figure}

Clearly, not all scalar power spectra can be extrapolated to the distortion scales
using a description in terms of a running spectral index. Projections can then be obtained
by performing forecasts for parameteric constraints on particular models. We exemplify
this on the curvaton scenario from Sec.~\ref{sec:curvaton}, using the
\Greens package~\cite{Chluba:2013vsa} for the MCMC analysis. As the fiducial model,
we take $a_\sigma=0$, $n_s^{(\rm inf)}=n_s(k_{\rm p})$ and
${\cal P}_{\zeta{\rm inf}}(k_{\rm p})={\cal P}_{\zeta}(k_{\rm p})$ are given by the Planck
central values from Eq.~(\ref{values:fiducial}), and $y_{\rm re}=4\times 10^{-7}$,
$\Delta^*=0$.
We define $\varrho=a_\sigma/a_{\rm inf}$
and show the projected bounds on $a_\sigma,\varrho$ in FIG.~\ref{FIG:MCMC:curvaton}.
We can conclude e.g. that PRISM may rule out a 60\% contamination
of the scalar perturbations from a curvaton with a blue spectrum at 95\% c.l.

\section{Conclusion} \label{sec:ccl}

We have performed the first exhaustive model-oriented analysis of the observational prospects for inflation in view of future observations of spectral distortions.  These are produced prior to recombination, when acoustic waves of Silk-damping scales dissipate their energy into the temperature monopole.   We find that only very few models can lead to distortions at a level that is detectable by the future PIXIE experiment.

After having introduced a methodology for the model selection, all 49 single-field inflation models listed in the \textit{Encyclopaedia Inflationaris} by Martin et al. in Ref.~\cite{Martin:2013tda}  have been considered.  We find that only one of them can satisfy simultaneously the Planck constraints on the primordial scalar power spectrum and generate a significant increase of power on smaller distortion scales, leading to a detectable level of intermediate and $\mu$-type distortions, while the $y$-type component should be dominated by other sources, such as the thermal SZ effect. This one scenario of interest is the original hybrid model with an instantaneous waterfall phase but super-Planckian field values.   The increase of power on small scales results from the field dynamics evolving from the field dominated towards the false vacuum dominated inflationary regime, where the spectrum is generically blue.   Finally, it is important to note that the region of parameter space  that is interesting in the present context appears strongly tuned. Therefore, it would be rather fortuitous, if distortions from single-field inflation were detected by PIXIE.   The situation could improve in the future in the context of a PRISM-class experiment, characterized by a ten times better sensitivity.  In this case, not only models with a sufficient increase of power on distortion scales could be tested but also models predicting a large running or, more generally, models with a substantial change of the amplitude of the scalar power spectrum between distortion and CMB anisotropy scales.  For our model selection, we find that some parameter ranges of the Non-Canonical K\"ahler and the Running Mass inflation models can be tested by a PRISM-like experiment. 

In addition, we have analyzed three effective scenarios of multi-field inflation:  a softly turning, a suddenly turning and a mild waterfall trajectory.  We find that softly turning trajectories do not generate any detectable distortions by PIXIE (they are nevertheless detectable by a PRISM-class experiment, but only for a very limited range of the parameters).  In the case of a sudden turn in the field space, a peak in the scalar power spectrum as well as damped oscillations are predicted on scales exiting the Hubble radius close to the turning point.  We find that an important and potentially detectable level of spectral distortions can be produced if the turning point is tuned to lie within the range of scales relevant for distortions, i.e. about  $8 \Mpc^{-1}\lesssim k \lesssim 10^4 \Mpc^{-1}$.   A peak in the scalar power spectrum is also a generic prediction of hybrid models ending with a mild waterfall.  The peak can be located in the range of scales that is of the present interest when inflation continues for about 40 e-folds of expansion during the waterfall.  This may lead to an important level of CMB distortions and we have determined the parameter range that will be 
probed with the PIXIE experiment.   It is important to note that the ratio between $\mu$-type and intermediate $i$-type distortions could be a good discriminator between the models.  

Finally, we have considered another possible scenario potentially leading to a strong distortion signal: a massive curvaton whose blue spectrum is subdominant on CMB angular scales and overtakes the inflaton spectrum on smaller distortion scales.   However, we find that the generated level of distortions does not reach the sensitivity PIXIE when we impose the bounds from ultra compact minihaloes.   But a not yet constrained part of the parameter space could be tested by an ultimate experiment like PRISM. 
 
Our analysis demonstrates that if $\mu$-type or intermediate $i$-type distortions are discovered by future experiments, they might be produced by only a few inflation models, in particular within multi-field scenarios, with the price being a somewhat tuned and contrived choice of parameters.  Therefore, one might then favour other potential sources of distortions like energy injection from decaying or annihilating particles or from primordial black hole evaporation.

\section*{Acknowledgements}

The authors warmly thank  Jens Chluba, Subodh Patil, Rishi Khatri, Juan Garc\'ia-Bellido and Christophe Ringeval for useful discussions and comments.  The work of SC is supported by the \textit{mandat de retour} program of the Belgian Science Policy (BELSPO). BG and YZ receive support from the Gottfried Wilhelm Leibniz programme
of the Deutsche Forschungsgemeinschaft (DFG) and from the DFG cluster of
excellence Origin and Structure of the Universe.

\bibliography{biblio}

\end{document}